\documentclass[preprint,12pt]{elsarticle}
\usepackage{amsmath}
\usepackage{amssymb}
\usepackage{multirow}
\usepackage{enumerate}
\usepackage{cases}
\usepackage{amsthm}
\usepackage{float}
\usepackage{booktabs}
\usepackage{makecell}
\usepackage{mathrsfs}
\usepackage{amsmath,amsfonts,amssymb,graphicx,mathtools,bm,tcolorbox,relsize}
\usepackage{color}
\usepackage{hyperref}
\usepackage{algorithm}
\usepackage{algpseudocode}

\usepackage[misc]{ifsym}
\usepackage{caption}

\usepackage{subfigure}
\usepackage{array}
\newcommand{\PreserveBackslash}[1]{\let\temp=\\#1\let\\=\temp}
\newcolumntype{C}[1]{>{\PreserveBackslash\centering}p{#1}}
\newcolumntype{R}[1]{>{\PreserveBackslash\raggedleft}p{#1}}
\newcolumntype{L}[1]{>{\PreserveBackslash\raggedright}p{#1}}

\makeatletter

\newcommand {\Rmnum} [1] {\expandafter \@slowromancap \romannumeral #1@}
\makeatother

\biboptions{compress}

\date{}

\begin{document}

\begin{frontmatter}

\title{Minimizing CNOT-count in quantum circuit of the extended Shor's algorithm for ECDLP}

\author{Xia Liu}	
\author{Huan Yang}
\author{Li Yang\corref{1}}
\ead{yangli@iie.ac.cn}
\cortext[1]{Corresponding author.}

\address{State Key Laboratory of Information Security, Institute of Information Engineering, CAS, Beijing, China}
\address{School of Cyber Security, University of Chinese Academy of Sciences, Beijing, China}
\address{Institute of Information Engineering, Chinese Academy of Sciences, Beijing, China}

\begin{abstract} 
Since the elliptic curve discrete logarithms problem (ECDLP) was proposed, it has been widely used in cryptosystem because of its strong security. Although the proposal of the extended Shor's algorithm offers hope for cracking ECDLP, it is debatable whether the algorithm can actually pose a threat in practice. From the perspective of the quantum circuit of the algorithm, we analyze the feasibility of cracking ECDLP with improved quantum circuits using an ion trap quantum computer.

We give precise quantum circuits for extended Shor’s algorithm to calculate discrete logarithms on elliptic curves over prime fields, including modulus subtraction, three different modulus multiplication, modulus inverse, and windowed arithmetic. Whereas previous studies mostly focused on minimizing the number of qubits or the depth of the circuit, we minimize the number of CNOTs, which greatly affects the time to run the algorithm on an ion trap quantum computer. First, we give the implementation of the basic arithmetic with the lowest known number of CNOTs and the construction of an improved modular inverse, point addition, and the windowing technique. Then, we precisely estimate the number of improved quantum circuits needed to perform the extended Shor's algorithm for factoring an $n$-bit integer, which is $1237n^3/\log n+2n^2+n$. We analyze the running time and feasibility of the extended Shor's algorithm on an ion trap quantum computer according to the number of CNOTs. Finally, we discussed the lower bound of the number of CNOTs needed to implement the extended Shor's algorithm.

\end{abstract}


\begin{keyword}
Elliptic curve discrete logarithm problem \sep
Extended Shor's algorithm \sep
Quantum circuits \sep
Ion trap quantum computer \sep
\end{keyword}

\end{frontmatter}
%




\section*{Introduction}

Elliptic curve cryptography (ECC) has attracted wide attention for its unique advantages since it was introduced in 1980s~\cite{miller1985use, koblitz1987elliptic}.The safety of ECC is based on the elliptic curve discrete logarithm problem(ECDLP), 
which is the discrete logarithm problem (DLP) on the cyclic subgroup with a point on the elliptic curve as the generator and more complex than DLP. Although there are many attempts to solve DLP, the best known classical algorithm for DLP is still exponentially complex~\cite{miyaji1992elliptic}. Fortunately, with the development of quantum computing, the emergence of quantum algorithms offers hope for solving such problems. The most representative and compelling quantum algorithm is Shor's algorithm~\cite{shor1994algorithms,shor1999polynomial}, which can theoretically solve DLP over multiplicative group for the prime fields
in polynomial time~\cite{shor1994algorithms, shor1999polynomial}.
This algorithm can be extended to elliptic curve groups (we call it extended Shor's algorithm in this paper), which makes ECDLP theoretically not difficult to a quantum computer, thus posing a threat to the cryptography system based on ECDLP. However, the gates number of a quantum algorithm's circuit determines the time to run the quantum algorithm on a quantum computer and the exact quantum gates number of the extended Shor's algorithm has not been analyzed. Therefore, it is debatable whether the extended Shor's algorithm can pose a threat to ECC, which is exactly what we are trying to do.

Quantum computers implement quantum computation by taking as input superposition quantum states representing all the different possible inputs and simultaneously evolving them into the corresponding outputs using a sequence of unitary transformations~\cite{yao1993quantum,nielsen2002quantum,chiribella2008quantum,dong2013quantum,nash2020quantum,kimura2021variational,liu2021cnot,liu2022mitigating,gao2022quantum}. Quantum computing can be described as a quantum circuit in which the unitary transformations are represented by quantum gates. The most basic quantum gates are control-NOT (i.e., CNOT) and single qubit gates. In an ion trap quantum computer, the operation time of the non-adjacent CNOT is much higher than that of other single quantum gates, and the CNOT can only operate serially~\cite{yang2013post}. Therefore, the number of CNOT contained in the quantum circuit of a quantum algorithm largely determines the running time of the quantum algorithm. 




Since the advent of the first quantum algorithm to attack ECC in 1995~\cite{boneh1995quantum}, the research in this field has attracted extensive attention.  Refs.~\cite{eicher1997using, proos2003shor, kaye2004optimized} proposed quantum algorithms that attack ECDLP defined on finite fields $F_p$ and $F_{2^m}$, respectively. Ref.\cite{roetteler2017quantum} studied the extended Shor's algorithm to attack ECDLP on $F_p$ and improved the algorithm of modular inversion in~\cite{proos2003shor}. The resources needed from the Toffoli gates point of view were $448n ^3\log_2(n)+ 4090n ^3$, but only $O(n^3)$ rough results were given for the CNOT gates. Ref.~\cite{haner2020improved} improved the Kaliski algorithm in the middle of~\cite{roetteler2017quantum}. Fewer T gates were used in the circuit of modular inversion and the windowed arithmetic in Ref.~\cite{gidney2019windowed} was briefly introduced to calculate ECDLP, but it is not discussed in detail. In view of the size of the quantum computer, i.e. the number of qubits, a quantum circuit for calculating the discrete logarithm problem on a binary elliptic curve is optimized in Ref.~\cite{banegas2021concrete}. 

Note that the resources required by the quantum circuit in previous papers did not analyse the number of CNOT gates in detail, but with the development of ion trap quantum computer, the number of CNOT gates largely determines the algorithm running time~\cite{knight1999ion}. Therefore, this paper analyzes the feasibility of the quantum algorithm to attack ECDLP by studying the number of CNOT gates in the circuit and discusses the application of windowed arithmetic in detail. It is worth noting that based on the physical limitations of quantum computers, we consider whether a sufficiently large quantum computer in the future can complete the extended Shor's algorithm in a reasonable running time, so we do not focus on the number of qubits.


\subsection*{Our contributions}
In this paper, we give precise quantum circuits for the extended Shor's algorithm to calculate discrete logarithms on elliptic curves over prime fields. More specifically, we have the following contributions.
\begin{enumerate}
    \item We construct and improve the circuits of basic operations including modulus subtraction, three different modulus multiplication, modulus inverse, windowed arithmetic and further improve the quantum circuit of extended Shor's algorithm.
    \item We combine window technique to focus on optimizing the number of CNOT gates, and further analyze the running time of extended Shor's algorithm on ion trap quantum computers according to the CNOT gates number we obtained. 
    \item We study whether the extended Shor's algorithm can be completed in a reasonable running time under the premise that the fault-tolerant quantum computer has enough space, further illustrating whether the Shor's algorithm can really pose a threat to cryptosystems such as ECC.
\end{enumerate}

\subsection*{Outline}
The rest of the paper is organized as follows. \nameref{sec: pre} section is the introduction of ECDLP and the elliptic curves group law. \nameref{sec: QC} section introduces the basic circuits to compute scalar multiplication on the elliptic curves groups required by the algorithm, including modular multiplication, modular inverse, and windowed arithmetic, etc.. In \nameref{sec: QC for ecdlp} section, we design a new method to calculate the point addition reversibly out-of-place (storing the results in a new register), which is different from the way of in-place (replacing the input value by the sum) in~\cite{roetteler2017quantum} and reduces the CNOT number. The \nameref{sec: dis} section is a discussion of the time required to attack ECDLP.

\section*{Preliminaries}
\label{sec: pre}
In this section, we first give a brief description of DLP, and then show the Shor's algorithm for solving DLP. Next, we elaborate on the algorithm for solving ECDLP, which we call the extended Shor's algorithm.
\subsection*{Shor's quantum algorithm for solving the DLP}
\textbf{DLP.} Let $g$ be a generator of a finite cyclic group $G$ with the known order ${\rm ord}(g)=k$, i.e $g^k=1$. The DLP over $G$ is defined as, given an element $x\in G$, determining the unique $r\in[0,|G|-1]$ such that $g^r=x$, then $r=\log_gx$.
Consider the case when G is the additive group $Z_N$, where $N$ is a positive integer and $\gcd(g,N)=1$. Here the DLP is to find $r$ satisfying $r\cdot g\equiv x\mod N$. The DLP over the $Z_N$ can be solved by finding the multiplicative inverse of $g$ modulo $N$ with the extended Euclidean algorithm in polynomial time $(O(\log_2^2N))$~\cite{proos2003shor}.
However, in group $G=Z_p^*$ (i.e., the multiplicative group modulo $p$ and $g^r\equiv x\mod p$), there was no classical algorithm to solve the DLP (i.e., calculate  
$r=\log_g x$) until Shor~\cite{shor1994algorithms, shor1999polynomial} in 1994 proposed a quantum algorithm that could theoretically solve this problem in polynomial time.




\textbf{Shor's quantum algorithm.} To be specific, the Shor's algorithm uses three quantum registers to solve the DLP, each quantum register has $n$ qubits and satisfies $p\leq q=2^n<2p$. The Shor's algorithm for DLP is shown as follows.

 \begin{algorithm}[H]
		\caption{\textbf{Shor's quantum algorithm for DLP}}
		\label{Alg:shor}
		\begin{algorithmic}[1]
			\Require 
		    $g, x$ and $p$, such that ${\rm gcd}(g, p)=1$.
			\Ensure  integer $r$ such that $g^r=x\mod p$.
			\State Prepare a $3(p-1)$-qubit initial state in three quantum registers $|0\rangle|0\rangle|0\rangle$.
			\State Apply the Hadamard transform $H^{\otimes 2(p-1)}$ to the first two quantum registers to obtain
            \begin{align}
                \frac{1}{p-1}\sum_{a=0}^{p-2}\sum_{b=0}^{p-2}|a\rangle|b\rangle|0\rangle.
            \end{align}
			\State Perform a unitary transformation $U_f$ such that $U_f|a\rangle|b\rangle|0\rangle\to|a\rangle|b\rangle|g^ax^{-b}\mod p\rangle$ to transform the initial state into
            \begin{align}
                \frac{1}{p-1}\sum_{a=0}^{p-2}\sum_{b=0}^{p-2}|a\rangle|b\rangle|g^ax^{-b}\mod p \rangle.
            \end{align}	
			\State Perform quantum Fourier transform on the first two registers to get the state 
            \begin{align}
                &\frac{1}{(p-1)q}\sum_{a=0}^{p-2}\sum_{b=0}^{p-2}\sum_{c=0}^{q-1}\sum_{d=0}^{q-1}\exp\left[\frac{2\pi i}{q}(ac+bd)\right]|c\rangle|d\rangle|g^ax^{-b}\mod p\rangle\nonumber\\
                =&\frac{1}{(p-1)q}\sum_{z=0}^{p-1}\sum_{c=0}^{q-1}\sum_{d=0}^{q-1}\sum_{g^{a-rb}=z\mod p}\exp\left[\frac{2\pi i}{q}(ac+bd)\right]|c\rangle|d\rangle|z\rangle.
            \end{align}
			\State Measure and obtain the probability of  $|c\rangle|d\rangle|z\rangle$ is:
            \begin{align}
                P_{c,d,z}=\frac{1}{(p-1)^2q^2}\left|\sum_{g^{a-rb}=z\mod p}\exp\left[\frac{2\pi i}{q}(ac+bd)\right]\right|^2,
            \end{align}
            determine $r$ with high probability by classical post-processing on the measured results.
		\end{algorithmic}
\end{algorithm}

Using the above algorithm, Shor proved that $r$ can be calculated with high probability in polynomial time. Based on the Shor's algorithm to solve DLP, next we show the case of ECDLP.

\subsection*{Extended Shor's quantum algorithm for solving the ECDLP}
\textbf{ECDLP.} Let $F_p$ be a field of characteristic $p\neq 2,3$. An elliptic curve over $F_p$ is the set of solutions $(x,y)\in F_p\times F_p$ to the equation 
\begin{align}
    y^2=x^3+Ax+B,\label{eq:ecc}
\end{align}
where $A,B\in F_p$ satisfy $4A^3+27B^2\neq0$,
together with the point $O$ at infinity.
The set of all the points on the elliptic curve is $E(F_p)=\{(x,y)|y^2=x^3+Ax+B;A,B\in F_p\}\cup\{\infty\}$. Then $E(F_p)$ forms Abelian group with a point addition operation and $O$ as the neutral element.
Let $P\in E(F_p)$ be a generator of $\langle P \rangle $, which is a cyclic subgroup of $E(F_p)$ of known order ${\rm ord}(P)=r$, i.e., $rP=O$. Similar to DLP, the goal of ECDLP is to find the unique integer $m\in \{1,...,r\}$ such that $mP=Q$, where $r,m\in F_p$ and $Q$ is a given point in $\langle P\rangle$. Hasse~\cite{hasse1936theorie} pointed out that the number of all the points on the elliptic curve is $\#E(F_p)=p+1-t,|t|\leq 2\sqrt{p}$. Thus the order of $\langle P\rangle$ is no larger than $p$. Therefore, when analyzing ECDLP on $\langle P\rangle$, the order can be set to $p$, which has no effect on the results.

\textbf{Extended Shor's quantum algorithm.} Different from the Shor's quantum algorithm, the extended Shor's algorithm uses two $n$-qubit and one $2n$-qubit registers with $n=\lceil\log_2p\rceil$ to solve the ECDLP. The specific algorithm for ECDLP is shown as follows.

 \begin{algorithm}[H]
		\caption{\textbf{Extended Shor's quantum algorithm for ECDLP}}
		\label{Alg:shor}
		\begin{algorithmic}[1]
			\Require 
		    An elliptic curve $E(F_p)$ with two points $P$ and $Q$.
			\Ensure  The smallest integer $m$ such that $Q=mP$.
			\State Prepare a $4n$-qubit initial states $|0\rangle^{\otimes n}|0\rangle^{\otimes n} |kP\rangle^{\otimes 2n}$ in three quantum registers.
			\State Perform a Hadamard transform on the first two registers to transform the initial state into 
            \begin{align}
                \frac{1}{2^n}\sum_{a=0}^{2^n-1}\sum_{b=0}^{2^n-1}|a\rangle|b\rangle|kP\rangle.
            \end{align}
			\State Perform a unitary transformation $U_f$ such that $U_f|a\rangle|b\rangle|kP\rangle \to |a\rangle|b\rangle|((a+k)P+bQ)\mod p\rangle$ to transform the initial state into
            \begin{align}
                \frac{1}{2^n}\sum_{a=0}^{2^n-1}\sum_{b=0}^{2^n-1}|a\rangle|b\rangle|((a+k)P+bQ)\mod p\rangle.
            \end{align}
			\State Perform a quantum Fourier transform on the first two registers to obtain the state
            \begin{align}
                \frac{1}{2^n}\sum_{a=0}^{2^n-1}\sum_{b=0}^{2^n-1}\sum_{c=0}^{2^n-1}\sum_{d=0}^{2^n-1}\exp[\frac{2\pi i}{2^n}(ac+bd)]|c\rangle|d\rangle|((a+k)P+bQ)\mod p\rangle.
            \end{align}
			\State Measure the first two registers and
            determine $m$ with high probability by classical post-processing on the measured results.
		\end{algorithmic}
\end{algorithm}

The initial state of the third register is $|kP\rangle$ instead of $|0\rangle$ to satisfy the point addition rule on the elliptic curve. Whether we use $|0\rangle$ or $|kP\rangle$ has no effect on the result of measuring probability. The detailed proof can be seen in the Appendix. 

\subsection*{Elliptic curves group law}
Before designing the circuits of the extended Shor's quantum algorithm, The elliptic curve group law on an affine Weierstrass curve
we give the law on the group of elliptic curves.

Let $P(x_1,y_1)\neq O, Q(x_2,y_2), R(x_3,y_3)\in\langle P\rangle , P+Q=R$, the elliptic curve group law on the Eq.~\eqref{eq:ecc} can be computed as follows:
\begin{equation}
		P+Q=R=\left\{
		\begin{aligned}
			&\qquad\qquad Q&,&\quad P=O,\\
			&\qquad\qquad O\quad(i.e \infty)&,&\quad P=-Q=(x_2,-y_2),\\
			&(\lambda^2-(x_1+x_2),\lambda(x_1-x_3)-y_1)&,&\quad others,
		\end{aligned}
		\right.
\end{equation} 
where $\lambda$ satisfies the following equation:
 \begin{equation}
 	\lambda=\left\{
 	\begin{aligned}
 		&\frac{y_2-y_1}{x_2-x_1}&,&\quad P\neq Q,\\
 		& \frac{3x_1^2+A}{2y_1}&,&\quad P=Q.
 	\end{aligned}
 	\right.
 \end{equation} 
Thus we have
 \begin{equation}
	2P=(x',y')=\left\{
	\begin{aligned}
		&((\frac{3x_1^2+A}{2y_1})^2-2x_1,\frac{3x_1^2+A}{2y_1}(x_1-x')-y_1)&,&\quad y_1\neq O,\\
		&\qquad \infty&,&\quad y_1=0.
	\end{aligned}
	\right.
\end{equation} 
Refs.~\cite{proos2003shor, roetteler2017quantum, haner2020improved} described the detailed steps of how to transform coordinates from $(x_1,y_1)$ to $(x_3,y_3)$. Since the purpose of this paper is to reduce the number of CNOT gates, we improve the previous steps of the coordinate transformation in \nameref{sec: QC for ecdlp} to reduce the number of CNOT gates but at a cost of increasing the number of qubits, which is not the focus of this paper.

\section*{Quantum circuits for algebraic problems}
\label{sec: QC}

In the implementation of the extended Shor's algorithm for ECDLP, the most important is to design a quantum circuit to compute scalar multiplication on the elliptic curves groups, i.e., $((a+k)P+bQ) \mod p$, which includes a series of modular operations. 
In this section, we design the circuits of modular subtraction and direct modular multiplication operations. Meanwhile, we improve a series of basic operations as well as including modular inversion and windowed arithmetic. In the following, the black triangle symbol in the circuits indicate that the corresponding qubit register is modified and holds the result of computation.

\subsection*{Modular subtraction}
Modular subtraction is divided into four parts: controlled and non-controlled constant modular subtraction, controlled and non-controlled quantum state modular subtraction. The difference between the constant and the quantum state is that the constant is known and can be ignored in quantum circuits.

Modular subtraction of a constant $|(x-y)\mod p\rangle $, where $y$ is a known constant and $p$ is the known $n$-bit constant, can be constructed by the following steps:
\begin{enumerate}
	\item Subtract $y$ from $|x\rangle$ to $|x-y\rangle$ using inverse circuit of the addition.
	\item If the highest bit of $|x-y\rangle$ is $1$ corresponding to $x-y<0$, then add $p$ to $|x-y\rangle$. Otherwise, do not operate.
	\item Compare the result of step $2$ with $(p-y)$. Uncompute the auxiliary bit and get $|(x-y)\mod p\rangle $.
\end{enumerate}

Next we give the detailed of the quantum circuit for performing addition and comparison.

(i) We use two circuits, $1-Add_y$ and $2-Add_y$, to perform addition, i.e., $|x\rangle |y\rangle |0\rangle \rightarrow |(x+y)_{0,...,n-1}\rangle |y\rangle |(x+y)_n\rangle$. The first two quantum registers both have $n$ qubits and the third one has $1$ qubit as the highest bit of the sum. The two circuits of addition are shown below.

\textcircled{1}: $1$-$Add_y$

\begin{figure}[h!]
    \centering
    \includegraphics[scale=0.5]{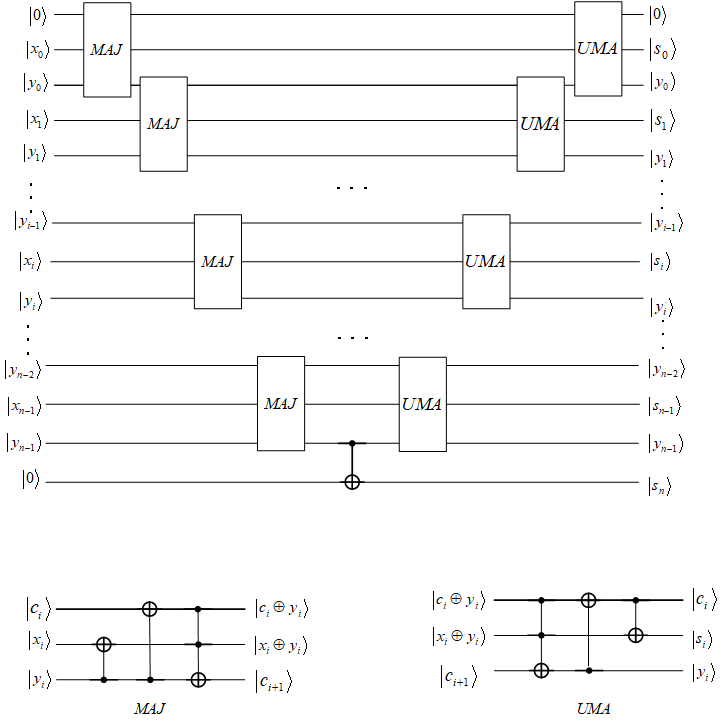}
    \caption{The first quantum circuit of addition $1$-$Add_y$ is constructed by $MAJ$ block and $UMA$ block. A $MAJ$ block and a $UMA$ block both have two CNOT gates and one Toffoli gate.}
    \label{fig:1-add}
\end{figure}

\begin{figure}[h!]
    \centering
    \includegraphics[scale=0.4]{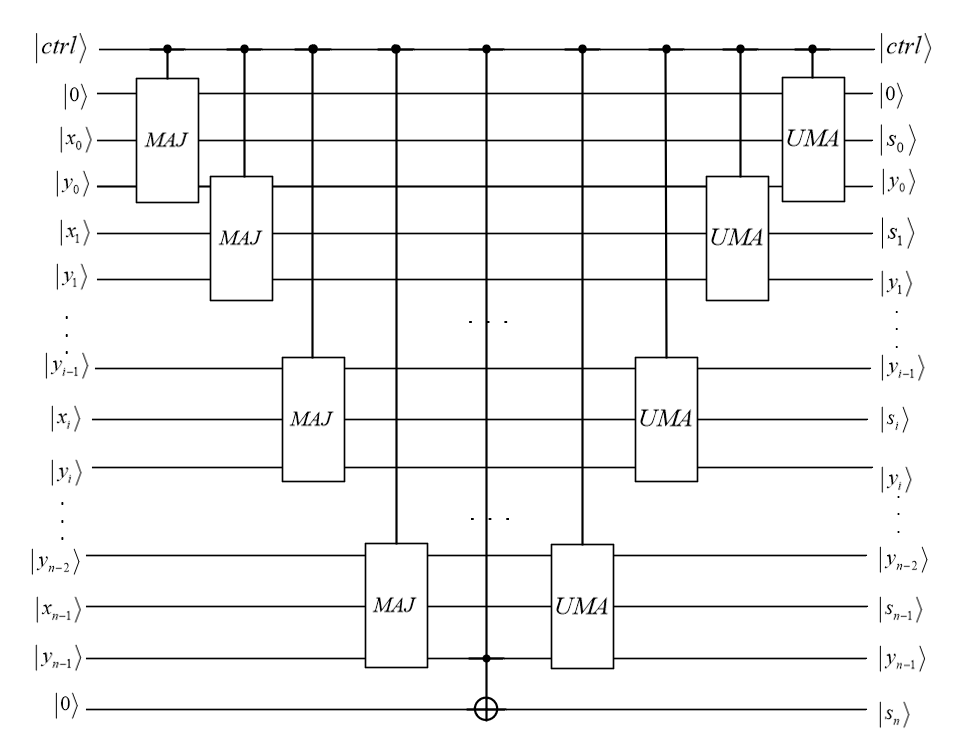}
    \includegraphics[scale=0.4]{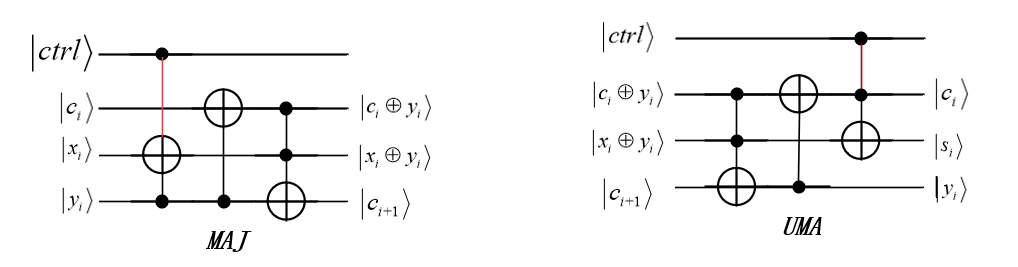}
    \caption{The controlled form of $1-Add_y$}
    \label{fig:ctrl-1-add}
\end{figure}

Ref.~\cite{cuccaro2004new} presented a way to calculate the addition as shown in Fig.~\ref{fig:1-add}, which we denote by $1$-$Add_y$. It shows that each $MAJ$ (i.e., compute the majority of three bits in place) and each $UMA$ (i.e., UnMajority and Add) contain two CNOT gates and one Toffoli gate. Thus, $n$-qubit $1-Add_y$ has a total $n$ $MAJ$s and $n$ $UMA$s, that is, $4n+1$ CNOT gates and $2n$ Toffoli gates. At the same time, based on the standard decomposition of the Toffoli gate into the Clifford+T set, we obtain that one Toffoli gate contains six CNOT gates~\cite{nielsen2002quantum}. Therefore, we conclude that the number of CNOT gates of an $n$-qubit $1-Add_y$ is $16n+1$. According to $1$-$Add_y$, we further design its controlled form in Fig.~\ref{fig:ctrl-1-add} with a CNOT number of $26n+6$.

\textcircled{2}: $2$-$Add_y$

\begin{figure}[H]
	\centering
	\includegraphics[scale=0.3]{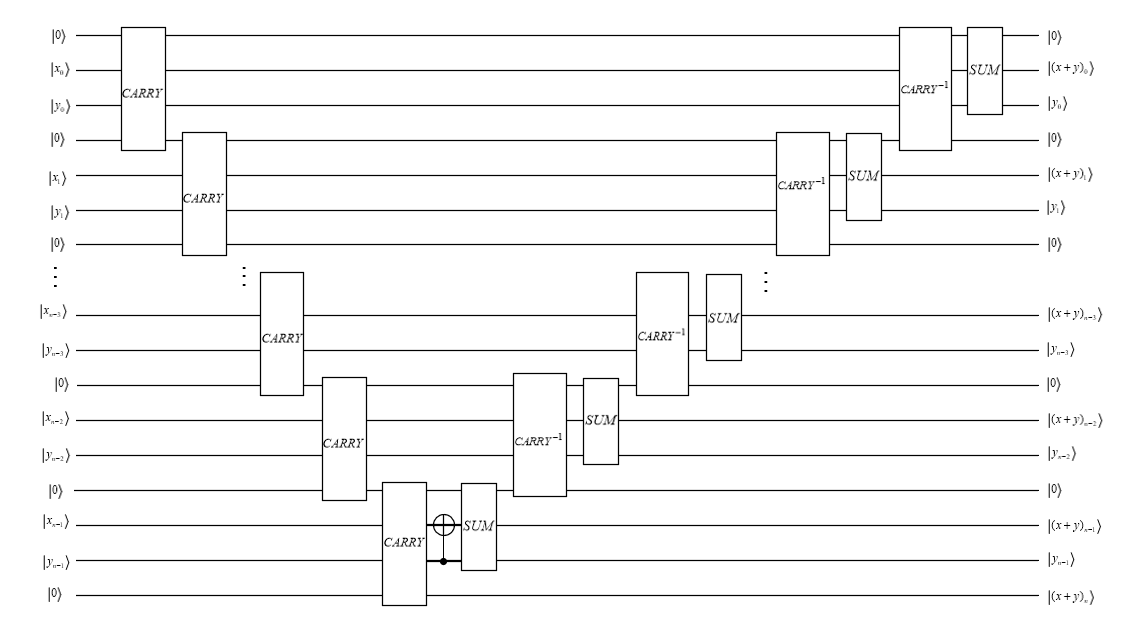}
	\caption{The second quantum circuits of addition 2-$Add_y$, where the blocks of $CARRY$ and $SUM$ are shown in the first line of Fig.~\ref{fig:2-add-constant}.}
	\label{fig:2-add}
\end{figure}

\begin{figure}[H]
    \centering
    \includegraphics[scale=0.45]{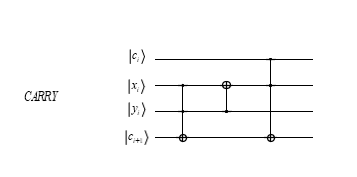}
    \includegraphics[scale=0.5]{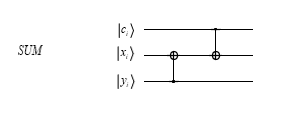}
     \includegraphics[scale=0.29]{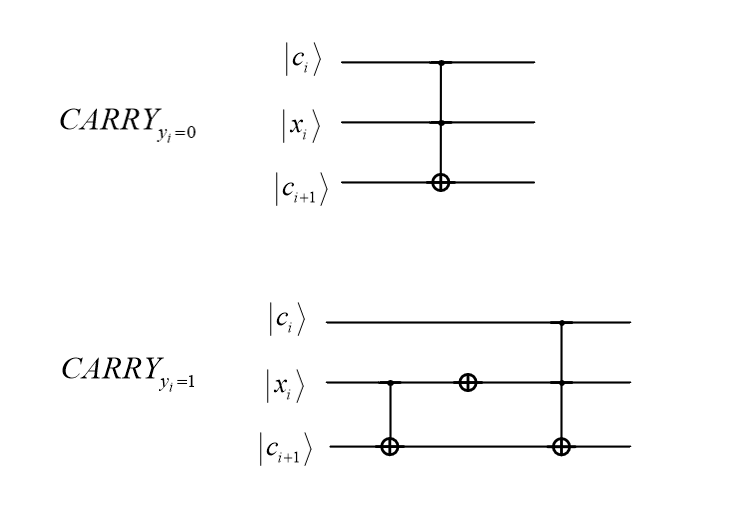}
     \includegraphics[scale=0.32]{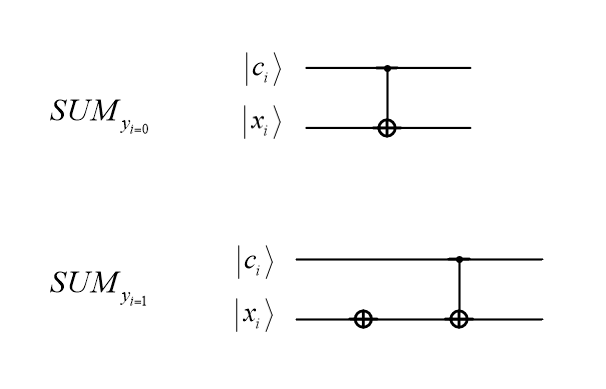}
    \caption{The form of $CARRY,SUM$ and the form of $CARRY$ and $SUM$ when y is known.}
    \label{fig:2-add-constant}
\end{figure}

Vedral et al.~\cite{vedral1996quantum} proposed another quantum circuit for calculating addition as shown in Fig.~\ref{fig:2-add}, where the blocks of $CARRY$ and $SUM$ are shown in Fig.~\ref{fig:2-add-constant} and the circuit of $CARRY^{-1}$ is the inverse order of the quantum gates in $CARRY$. When the addend $y$ is known, Markov et al.~\cite{markov2012constant} modified the $CARRY,SUM$ to the form shown in the last two rows of Fig.~\ref{fig:2-add-constant}, that is, the $y$ is omitted. At this point, one $CARRY$ (or $CARRY^{-1}$) contains on average $1$ Toffoli gate and $\frac{1}{2}$ CNOT, and one $SUM$ has on average $1$ CNOT. Therefore, an $n$-qubit $2$-$ADD_y$ has a total $n$ $CARRY$s, $n$ $SUM$s, $n-1$ $CARRY^{-1}$ and $1$ additional CNOT. Combine with six CNOT of one Toffoli, we conclude that the number of CNOT in $2-Add_y$ is $14n-5.5$ when y is known.

\begin{figure}[H]
    \centering
	\includegraphics[scale=0.6]{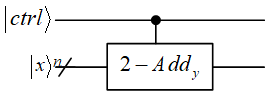}
 \qquad
	\includegraphics[scale=0.55]{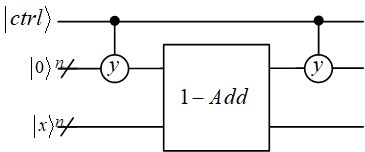} 
    \caption{The original controlled form of $2$-$Add_y$ and the new controlled form of $2$-$Add_y$ when y is known.}
    \label{fig:ctrl-2-add}
\end{figure}

The left circuit in Fig.\ref{fig:ctrl-2-add} is a common controlled form of $2$-$Add_y$, while the right one proposed by Ref.~\cite{haner2020improved} gives a simpler controlled form. That is, the control bits $ctrl$ use NOT gates to control the known addend $y$ to store in an $n$-qubit auxiliary register, then the addend $y$ cannot be omitted and need to use $1$-$Add$ to sum. Finally, repeat the storage operation to restore the auxiliary bits. Since encoding a known $n$-qubit addend $y$ into the circuit requires an average of $\frac{1}{2}$ CNOTs, combined with the $1-Add$, we conclude that $2$-$Add$ requires $17n+1$ CNOTs.

(ii) Now we also use two circuits, $1$-$Comp_y$ and $2$-$Comp_y$, to perform comparison. Compare $x$ and $y$ by whether the highest bit of $x-y$ is $0$ or $1$. When the highest bit is $0$ then $x-y>0$, otherwise $x-y<0$. The difference between these two circuits is that $1$-$Comp_y$ applies to the case where $y$ is a known constant, while $2$-$Comp_y$ can be used either for $y$ known or for $y$ unknown. Although we use $2$-$Comp_y$ for all the comparisons covered in this paper, both circuits are presented for the sake of the completeness of the method. Details of the two circuits are shown below.

\textcircled{1}: $1$-$Comp_y$

\begin{figure}[H]
	\centering
	\includegraphics[scale=0.3]{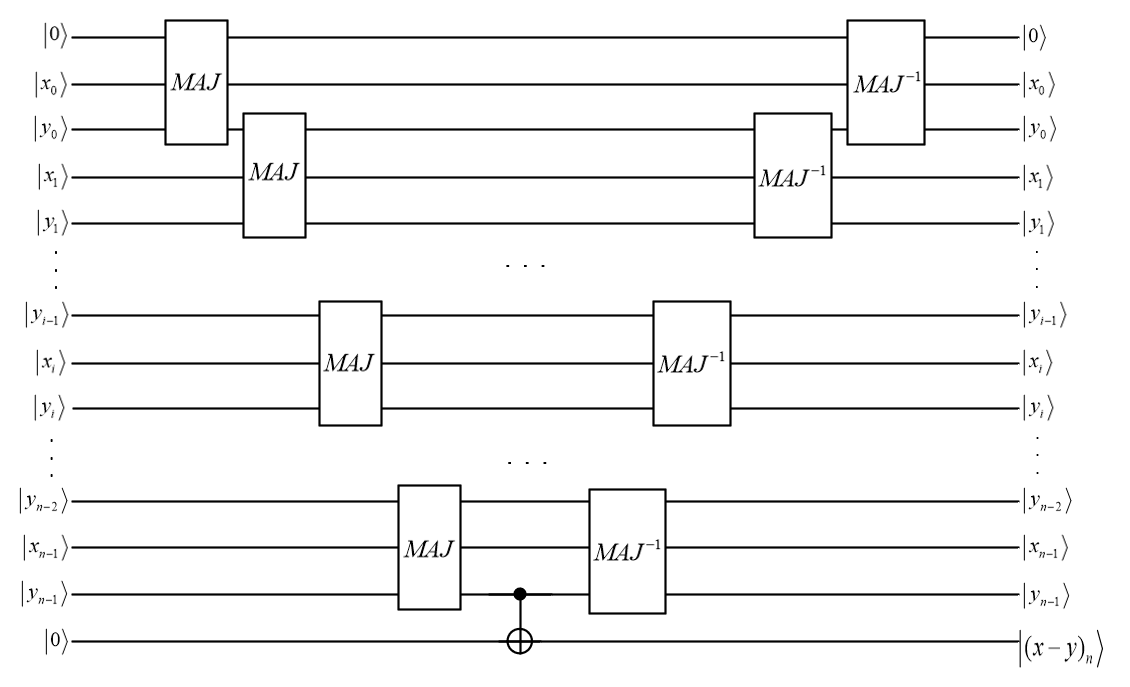}
	\caption{$1-Comp_y$. The $y$ is a known constant.}
	\label{fig:1-comp}
\end{figure}

\begin{figure}[H]
	\centering
	\includegraphics[scale=0.3]{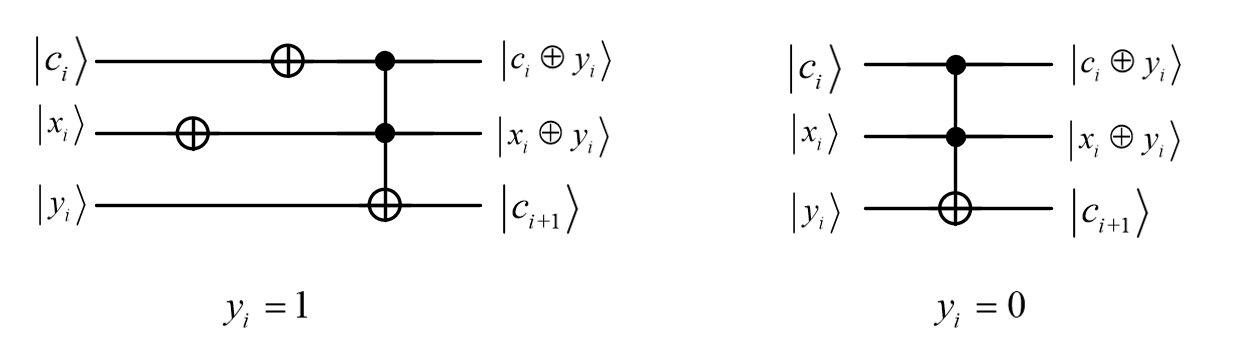}
	\caption{The form of $MAJ$ when y is known}
	\label{fig:1-comp-y-know}
\end{figure}

$1$-$Comp_y$ in Fig.~\ref{fig:1-comp} is obtained by modifying $1$-$Add_y$ so that it output only the highest bit of $|x-y\rangle$~\cite{markov2012constant}. But the premise is that the input is $-y+2^n$ instead of $y$, which means this way only works if $y$ is a known constant instead of an unknown quantum state.
When $y$ is known, the $MAJ$ can be simplified to Fig.~\ref{fig:1-comp-y-know}, that is, one $MAJ$ contains $1$ Toffoli. Thus the number of CNOT in $1-Comp_y$ is $12n+1$. 

\textcircled{2}: $2$-$Comp_y$

\begin{figure}[H]
	\centering
	\includegraphics[scale=0.42]{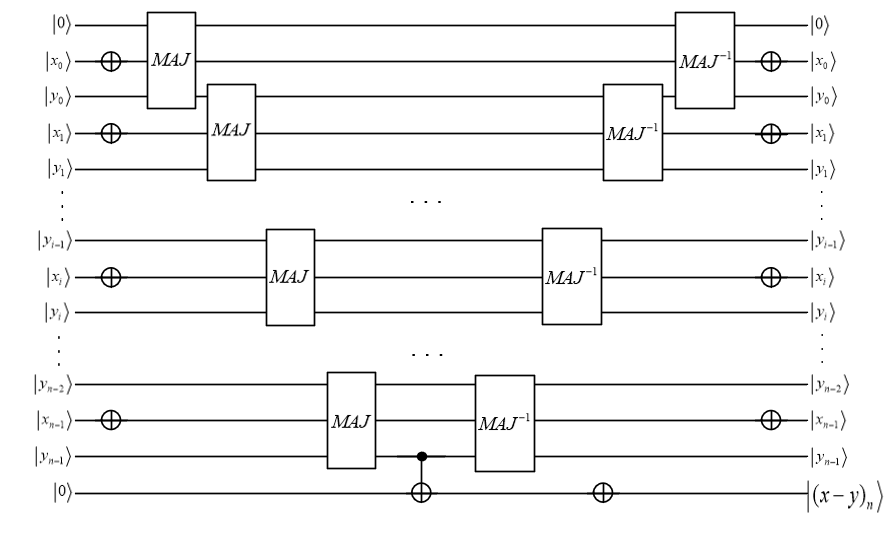}
	\caption{2-$Comp_y$. The $y$ can be either known or unknown.}
	\label{fig:2-comp}
\end{figure}

\begin{figure}[H]
	\centering
	\includegraphics[scale=0.32]{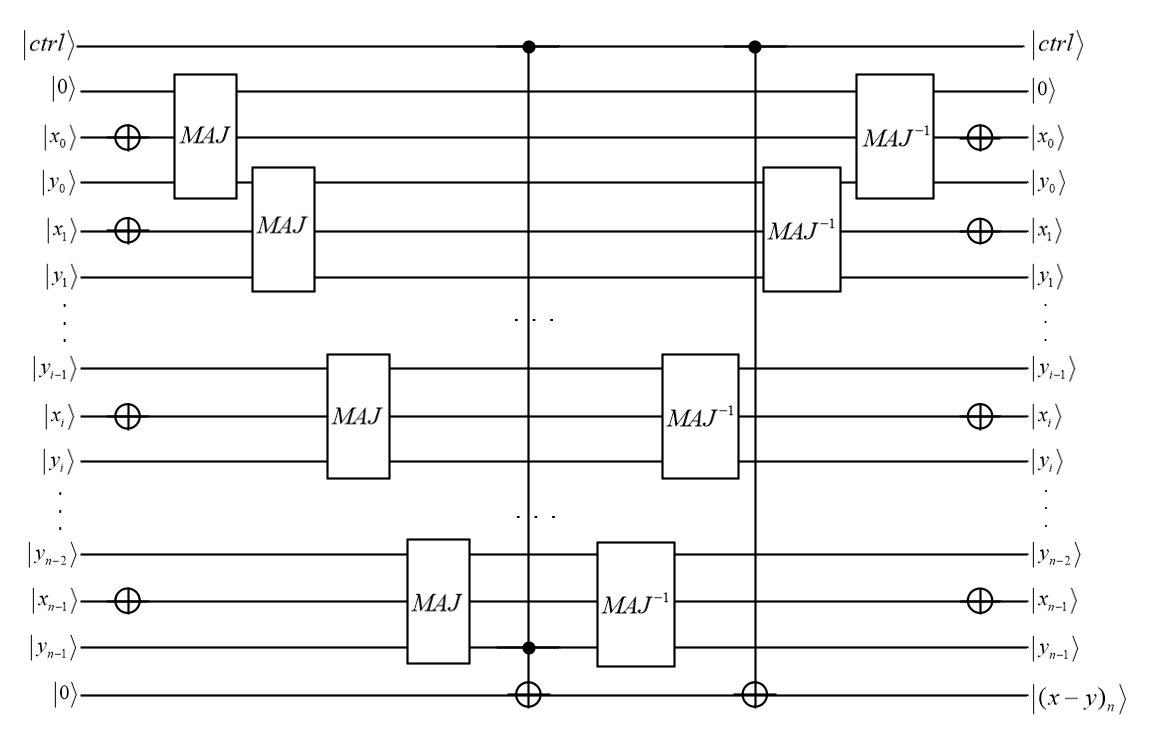}
	\caption{The controlled form of 2-$Comp_y$.}
	\label{fig:2-comp-ctrl}
\end{figure}

Although the 1-$Comp_y$ does not work when the minus $y$ is an unknown quantum state, it can be modified to not precompute $-y+2^n$. The specific steps are as follows. Firstly, input $x,y$ and flip each of the $x$ bits to get $2^n-1-x$. Then use 1-$Comp_y$ to get the highest bit of $2^n-1-x+y$, the flip of $x_{0,...,n-1}$ and $y_{0,...,n-1}$. Finally, flip each of the $x$ bits and the highest bit of $2^n-1-x+y$ to recover $x$ and get $(x-y)_n$ which means the the highest bit of $x-y$. Following above steps to obtain 2-$Comp_y$ as shown in Fig.~\ref{fig:2-comp}. We see that $2$-$Comp_y$ not only applies to where $y$ is a known constant but also applies to an unknown quantum state. The number of CNOT in the former is $12n+1$, which is the same as 1-$comp_y$, and the latter is $16n+1$. Fig.~\ref{fig:2-comp-ctrl} is the controlled form of 2-$Comp_y$. The corresponding CNOT numbers are $12n+7$ and $16n+7$, respectively.

\begin{figure}[H]
	\centering
	\includegraphics[scale=0.55]{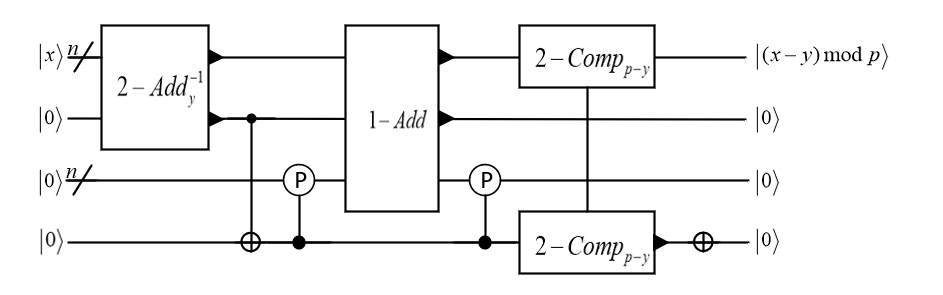}
	\caption{Circuit of the constant modular subtraction $ModAdd^{-1}(\cdot)$. Since addition is the inverse operation of subtraction and $y$ is a known constant, we can use the reversible circuit of $2$-$Add_y$ to compute $|x-y\rangle$ and denote it $2$-$Add_y^{-1}$.}
	\label{fig:modsub}
\end{figure}

After showing the circuits of addition and comparison in steps (i) and (ii), we design the constant modular subtraction circuit $ModAdd^{-1}(\cdot)$ in Fig.~\ref{fig:modsub}, which contains one $2-Add_y^{-1}$, one CNOT, one $1-Add$, one $2-Comp_{p-y}$ with the known constant $p-y$, and two circuits of encoding $p$. Thus we conclude that the number of CNOT in $ModAdd^{-1}(\cdot)$ is $43n-2.5$. Calculate $|(x+y)\mod p\rangle$ using the reversible circuit of $ModAdd^{-1}(\cdot)$, which denotes $ModAdd(\cdot)$.

\begin{figure}[H]
	\centering
	\includegraphics[scale=0.55]{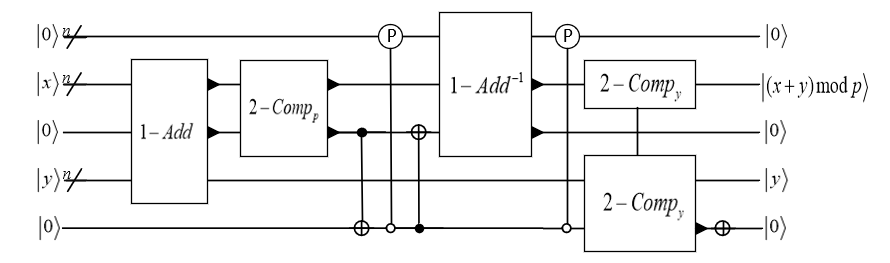}
	\caption{Circuit of the quantum state modular addition $ModAdd$. Since addition is the inverse operation of subtraction, we can use the reversible circuit of $1$-$Add$ to compute $|x-y\rangle$ and denote it $1$-$Add^{-1}$.}
	\label{fig:modadd}
\end{figure}

The quantum state modular addition circuit $ModAdd$ can be obtained in a similar way, which is shown in Fig.~\ref{fig:modadd}. Different from the constant modular addition, $ModAdd$ contains one $1-Add$, two $2-Comp_y$ with the known constant $y$, two CNOTs, one $1-Add^{-1}$, and two circuits of encoding $p$. Thus, we conclude that the number of CNOTs in $ModAdd$ is $61n+6$. Furthermore, using the reversible circuit $ModAdd$ we can calculate quantum state modular subtraction. 

\begin{figure}[H]
	\centering
	\includegraphics[scale=0.5]{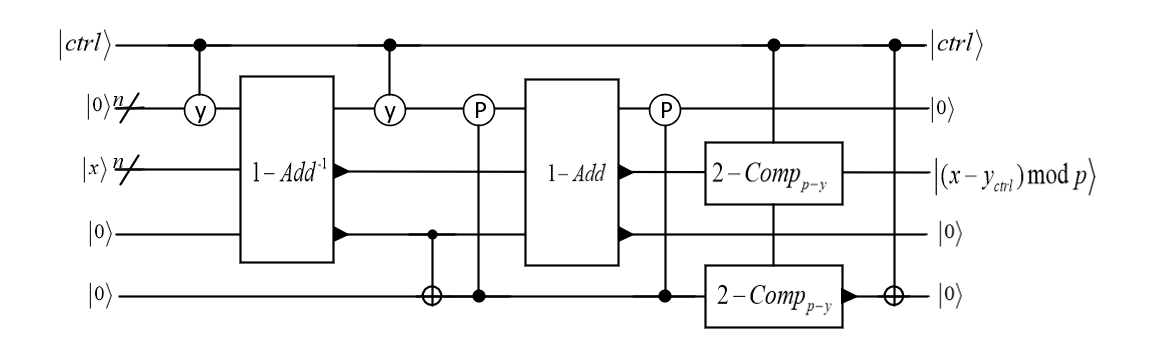}
	\caption{The controlled form of $ModAdd^{-1}(\cdot)$.}
	\label{fig:ctrl-modsub}
\end{figure}

\begin{figure}[H]
	\centering
	\includegraphics[scale=0.5]{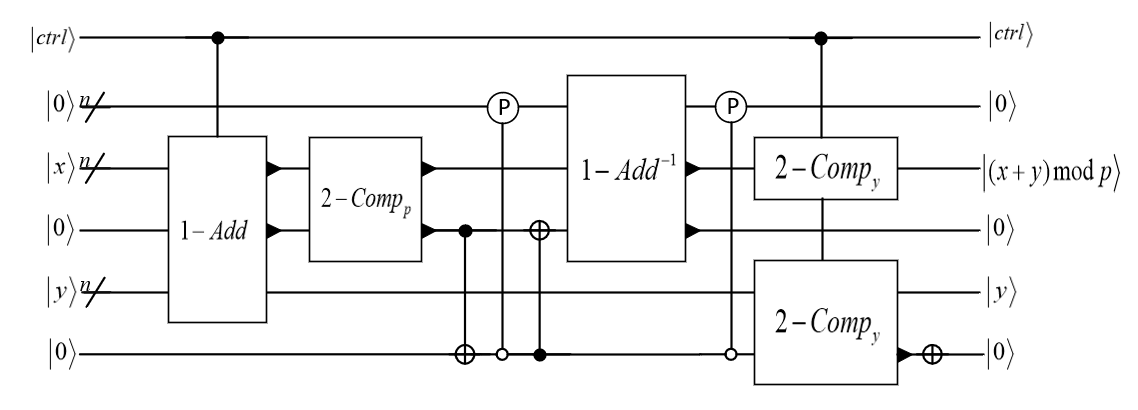}
	\caption{The controlled form of $ModAdd$.}
	\label{fig:ctrl-modadd}
\end{figure}

The controlled form of $ModAdd^{-1}(\cdot)$ and $ModAdd$ are shown in Fig.~\ref{fig:ctrl-modsub} and Fig.~\ref{fig:ctrl-modadd}, respectively. The corresponding CNOTs numbers are $46n+11$ and $71n+17$, respectively. 

\subsection*{Negation} 
\begin{figure}[H]
    \centering
	\includegraphics[scale=0.3]{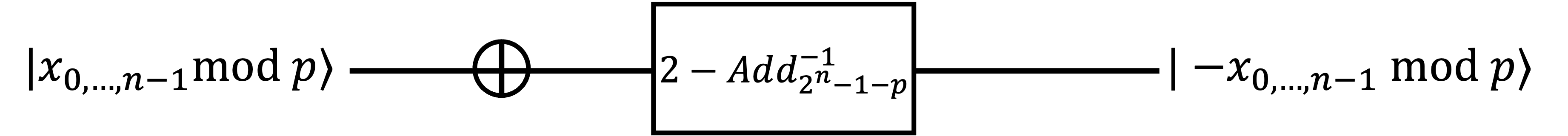}
    \caption{The circuit of negation $NegMod$.}
    \label{fig:neg}
\end{figure}

\begin{figure}[H]
    \centering
	\includegraphics[scale=0.4]{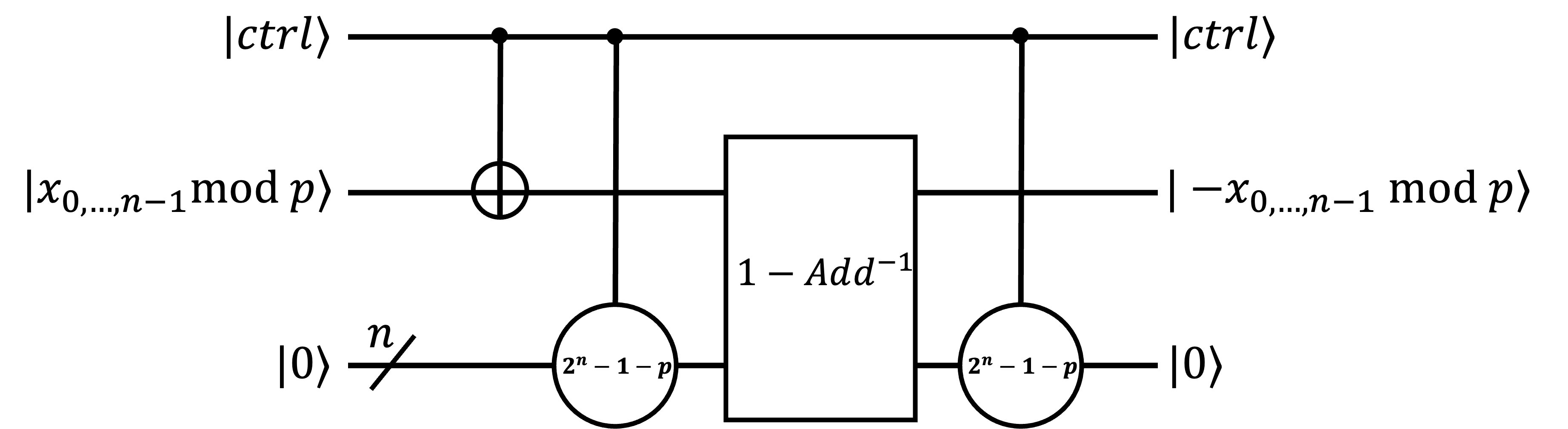}
    \caption{The controlled form of $NegMod$.}
    \label{fig:ctrl-neg}
\end{figure}

Given the value of $x\mod p$, it is easy to calculate $-x\mod p$ algebraically. Conversely, performing this calculation using quantum circuit is difficult.
In order to solve this problem, Markov et al.~\cite{markov2012constant} indicated that it can be done by first flipping each of the bits $x$ to get $(2^n-1-x)$, then subtracting $(2^n-1-p)$ from $2-Add^{-1}$ to get the result. According to these two steps, Fig.~\ref{fig:neg} shows the circuit of calculating $-x\mod p$ and Fig.~\ref{fig:ctrl-neg} is its controlled form. The number of CNOT is equal to $2-Add^{-1}$ in $NegMod$, i.e. $14n-5.5$, while in controlled circuit is $18n+1$.

\subsection*{Modular shift}
For constructing the circuit of modular shift, i.e., $|x \mod p \rangle\rightarrow|2x\mod p\rangle$, we first show the circuits of the binary shift. The functions of the binary shift are as follows.
\begin{align}
    \text{Left shift $l-shift$: } &|0x_{n-1}\cdots x_1x_0\rangle\longrightarrow|x_{n-1}\cdots x_1x_00\rangle;\nonumber\\
    \text{Right shift $r-shift$: } &|x_{n-1}\cdots x_1x_00\rangle\longrightarrow|0x_{n-1}\cdots x_1x_0\rangle\nonumber.    
\end{align}
Before the method to implement shift is to use SWAP gate, that is, the second method below. However, we note that there is no need to swap two qubits with a SWAP operation if a qubit is known to be in the state of $|0\rangle$. Hence, we reconstruct the modular shift circuit for an $n$-qubit quantum register to reduce the CNOT number, that is, the first method below.

\begin{figure}[H]
	\centering
	\includegraphics[scale=0.6]{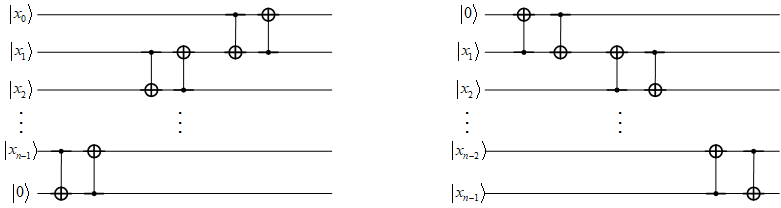}
	\caption{First way to perform binary shift, $l-shift$ and $r-shift$, respectively.}
	\label{fig:1-shift}
\end{figure}

\textcircled{1} The first shift method shown in Fig.~\ref{fig:1-shift} requires $2n$ CNOT, and the controlled form uses one control qubit to control each CNOT, which needs $2n$ Toffoli gates. Thus, the controlled form needs $12n$ CNOT.

\begin{figure}[H]
	\centering
	\includegraphics[scale=0.6]{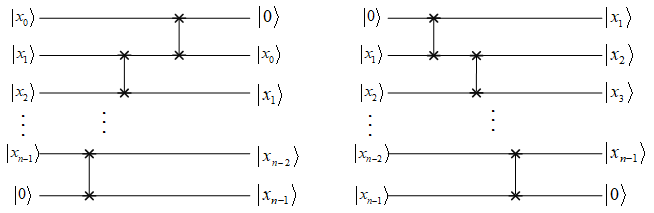}\\
	\includegraphics[scale=0.65]{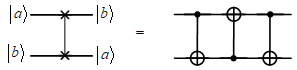}
	\caption{Second way to perform binary shift, $l-shift$ and $r-shift$, respectively.}
	\label{fig:2-shift}
\end{figure}

\textcircled{2} The second shift method shown in Fig.~\ref{fig:2-shift} requires $3n$ CNOT. Different from the first method, the controlled form of second method just needs to use one qubit to control the middle CNOT in each SWAP gate. Then the circuit requires $2n$ CNOTs and $n$ Toffoli gates in total, that is, $8n$ CNOT.

Based on the above two method to perform modular shift, we can choose an appropriate circuit to minimize the number of CNOT gates, that is, choose the second way when a controlled mode is involved, otherwise, choose the first method.

\begin{figure}[H]
	\centering
	\includegraphics[scale=0.4]{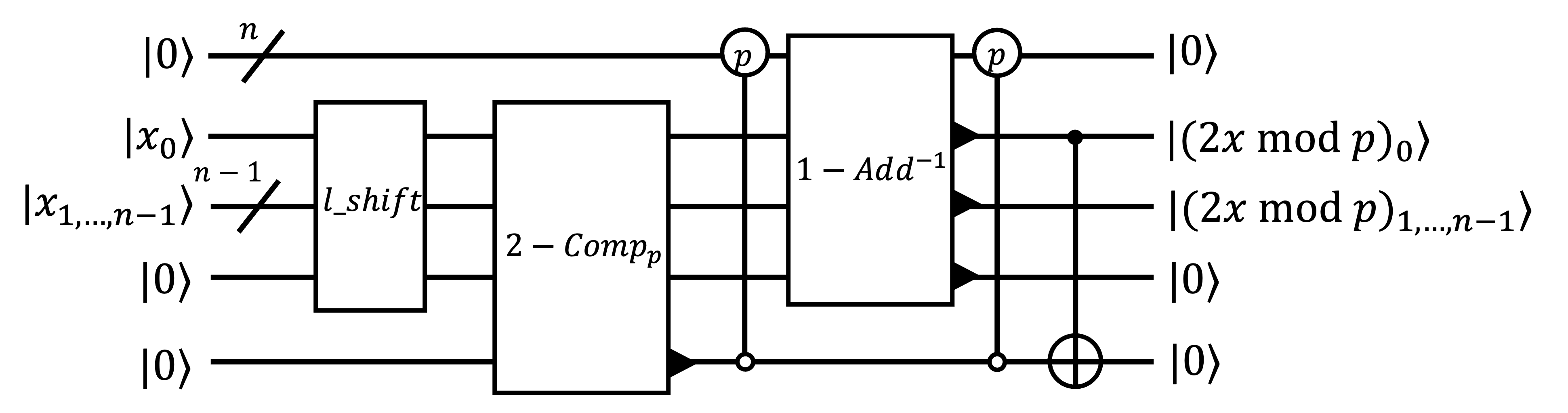}
	\caption{Circuit of the modular shift, $ShiftMod$.}
	\label{fig:shiftmod}
\end{figure}

As shown in Figure~\ref{fig:shiftmod}, we improve the modular shift, by replacing the subtraction of the constant $p$ with a comparison of the constant $p$. The CNOT count of our modular doubling is $31n+15$ by selecting the appropriate binary shift method.

\subsection*{Modular multiplication}
There are three kinds of modular multiplication methods: fast modular multiplication, Montgomery modular multiplication and direct modular multiplication. The first way is to compute by repeating modular and conditional modular additions. The second way is often the most efficient choice for modular multiplication when modulus $p$ is not close to a power of $2$. The last method is to calculate it in the most direct way, that is, first do the binary multiplication and then subtract multiples of $p$.

\textbf{Fast modular multiplication.} In Ref.~\cite{proos2003shor}, the fast modular multiplication is used to calculate the modular multiplication and the circuit of this method is designed in detail in the section 3.3 of Ref.~\cite{roetteler2017quantum}, which requires $104n^2-86.5n-11.5$ CNOTs. Furthermore, module addition and module shift in the fast modular multiplication apply the circuits mentioned earlier in this paper.

\textbf{Montgomery modular multiplication.} According to the Montgomery algorithm~\cite{kaliski1995montgomery}, input $x,y$, we can get $(x\cdot y\cdot 2^{-n}\mod p)$, where $2^{n-1}<p<2^n$ and Roetteler et al.~\cite{roetteler2017quantum} gave a specific quantum circuit. Combined with the basic arithmetic circuit improved before in this paper, appropriate circuits are selected to obtain the following Montgomery modular quantum circuit, the result in $M$-$Mul$ is $(x\cdot y\cdot 2^{-n}\mod p)$, where the $Add$ is 1-$Add$, $Add^{-1}$ is the constant subtraction 2-$Add^{-1}$. The inverse operation $M$-$Mul^{-1}$ is used to restore the auxiliary bits. The entire quantum circuit of Montgomery modular multiplication is a combination of $M$-$Mul$ and $M$-$Mul^{-1}$ with a CNOT number of $90n^2+78n-9$. Actually, to obtain the value $(x\cdot y\cdot 2^{-n}\mod p)$, we still need to set $n$ CNOTs to encode the value into an extra $n$-qubit auxiliary bits before performing $M$-$Mul^{-1}$.
\begin{figure}[H]
	\centering
	\includegraphics[scale=0.48]{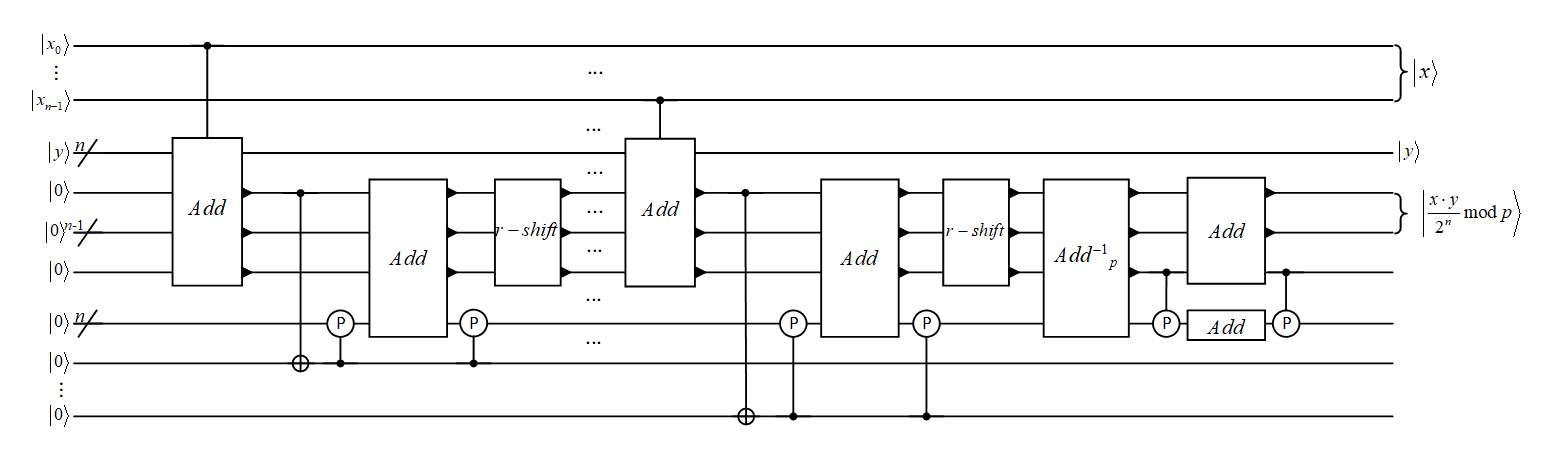}
	\caption{The partial quantum circuit of Montgomery modular multiplication $M$-$Mul:|x\rangle |y\rangle |0\rangle \rightarrow|x\rangle |y\rangle |x\cdot y\cdot2^{-1}\mod p\rangle $ }
	\label{fig:M-MM}
\end{figure}

\textbf{Direct modular multiplication.} Now we give a method to construct the circuit of modular multiplication according to its calculation. The main idea is that $x\cdot y=(kp+x\cdot y)\mod p, k=\lfloor\frac{x\cdot y}{p}\rfloor$, where $1<x,y<p$, so $x\cdot y<p^2<2^np$, thus 
\begin{align*}
x\cdot y\mod p=&x\cdot y-kp\\
=&\sum_{i=0}^{n-1}2^ix_i\cdot y-\sum_{i=0}^{n-1}2^ik_ip,\qquad k_i\in\{0,1\}.	
\end{align*}
Then by comparing the sizes of $x\cdot y$ and $2^ip$ to obtain the target result. Since this method is constructed directly according to the calculation, we call it direct modular multiplication.
More specifically, it is divided into the following four steps.
\begin{enumerate}
	\item Calculate the value of $x\cdot y$.
	\item Calculate the value of $x\cdot y-2^ip(i=n-1,...,0)$, i.e., $(xy)_i...(xy)_{n+i}-p$. If the highest bit of the result is 1, then add $p$ to the result.
	\item Repeat step $2$ until $i=0$;
	\item Repeat the reverse of steps $2$ $1$ in sequence to recover the auxiliary qubits.
\end{enumerate}

\begin{figure}[H]
	\centering
	\includegraphics[scale=0.4]{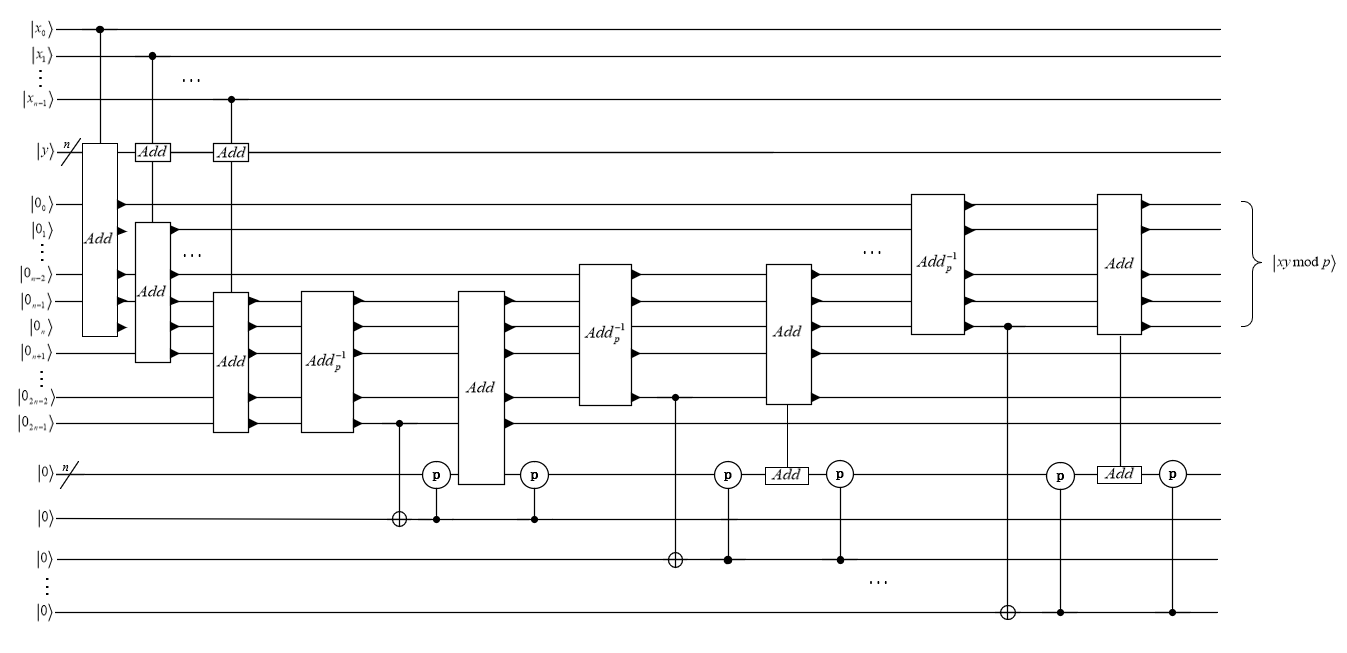}	
	\caption{The partial quantum circuit of direct modular multiplication $D$-$Mul:|x\rangle|y\rangle|0\rangle\rightarrow|x\rangle|y\rangle|x\cdot y\mod p\rangle$.}
	\label{fig:D-MM}
\end{figure}

According to the first three steps, we can obtain the following partial quantum circuit $D$-$Mul$. The circuit of step $4$ to restore the auxiliary bit is denotes $D$-$Mul^{-1}$, i.e. the inverse operation of $D$-$Mul$, where the $Add$ and $Add_p$ are 1-$Add$ and 2-$Add$ respectively in Fig.~\ref{fig:D-MM}. Thus the whole quantum circuit of direct modular multiplication needs $114n^2+5n$ CNOTs. Similar to the Montgomery modular multiplication, to obtain the value $(x\cdot y\mod p)$ we still need to set $n$ CNOTs to encode the value into an extra $n$-qubit auxiliary bits before performing $D$-$Mul^{-1}$.

\subsection*{Modular inversion}
The most common method of modular inverse is the extended Euclidean algorithm (EEA). Proos et al.~\cite{proos2003shor} described the idea of using EEA to calculate modular inverse and it required $O(n)$ times of division in total and each step was performed $O(n^2)$ times. However, implementing the EEA in a quantum circuit is very complicated, then we consider to use Montgomery inversion algorithm described in detail in Ref.~\cite{roetteler2017quantum} and they repeated the Montgomery-Kaliski round function $2n$ times to get $x^{-1}R\mod p$. Subsequently, Haner et al.~\cite{haner2020improved} improved this algorithm and its circuit used fewer CNOTs, but used the same modular inversion circuit. In this paper, we choose the improved algorithm in Ref.~\cite{haner2020improved} as the round function and redesign a simpler circuit to calculate modular inversion.

For inputs $x,p,n,p>x>0,2^{n-1}<x<2{n}$, the Montgomery-Kaliski algorithm consists of two steps. First, calculate gcd$(x,p)$ and $x^{-1}\cdot2^{k}\mod p$. Second, calculate $x^{-1}\cdot 2^n\mod p$. When the input quantum state is a superposition state, the number of iterations $k$ in the first step is related to the integer $x$ corresponding to a certain ground state. Considering all possible ground states in the superposition state, the first step requires $2n$ rounds of iteration. However, before each round, it is necessary to judge whether the iteration process in the corresponding ground state has ended by determining whether $v$ is 0, so as to determine whether this round is really iterated. Due to $k>n$, all ground states of the input superposition state need to go through the first $n$ rounds of iteration and only need to judge whether $v$ is 0 before the iteration of the last $n$ rounds. In the second step, the intermediate result $x^{-1}\cdot 2^k\mod p$ is shifted to the right by $k-n$ bits. In the last $n$-round iteration of the first step, the results of the subsequent ShiftMod of the second step is stored in the auxiliary qubit and $x^{-1}\cdot2^n\mod p$ is obtained.

\begin{figure}[H]
	\centering
	\includegraphics[scale=0.38]{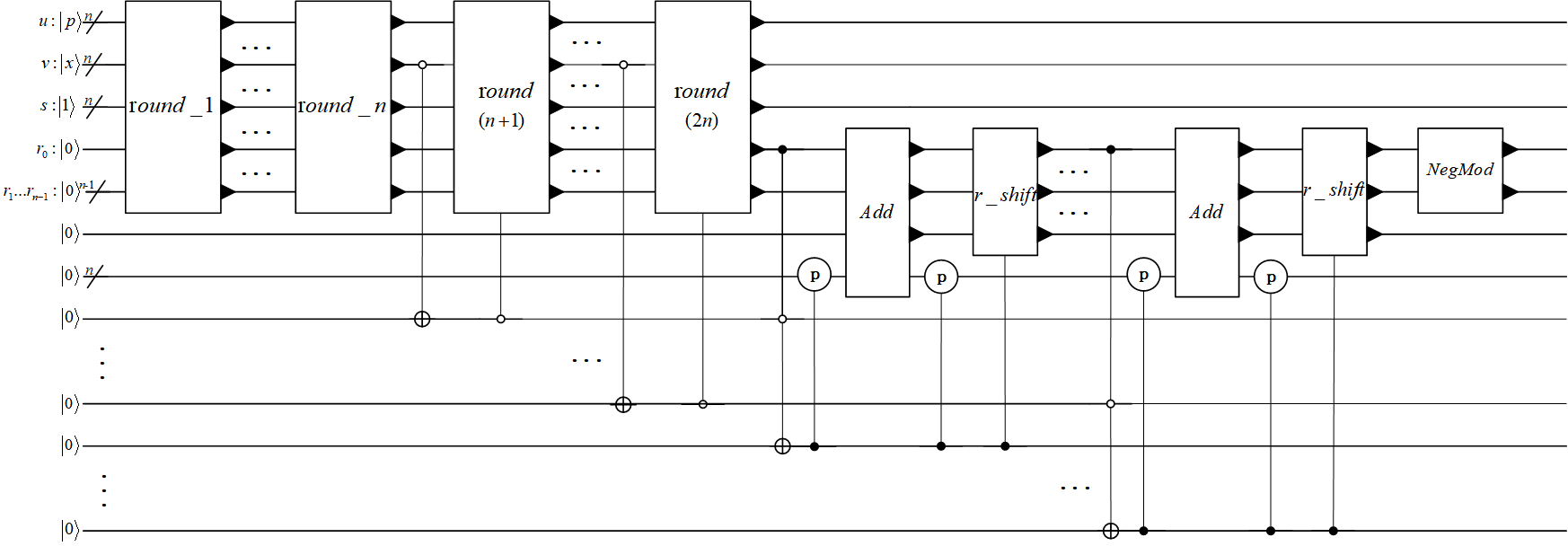}
	\caption{The partial quantum circuit of modular inversion $Inv:|x \mod p\rangle|0\rangle \rightarrow|x\rangle|x^{-1}\cdot2^n\mod p\rangle$.}
	\label{fig:MI}
\end{figure}

Combining the round function circuit in Figure 6 (b) of Ref.~\cite{haner2020improved} with the above algorithm steps, the quantum circuit of the modular inversion $Inv$ in Fig.~\ref{fig:MI} is obtained. The quantum circuit for restoring the auxiliary bits is $Inv^{-1}$, i.e. the inverse operation of $Inv$ and the complete quantum circuit is a combination of $Inv$ and $Inv^{-1}$.

According to $Inv$, we can obtain the whole quantum circuit of modular inversion needs $578n^2+283n-13$ CNOTs.
\subsection*{Windowed arithmetic}
In this section, we use the window form described in Ref.~\cite{gidney2019windowed} to design the quantum circuit to attack ECDLP, reducing the CNOT number $N$ from $O(n^3)$ to $O(n^2)<N<O(n^3)$.

\begin{figure}[H]
	\centering
	\includegraphics[scale=0.6]{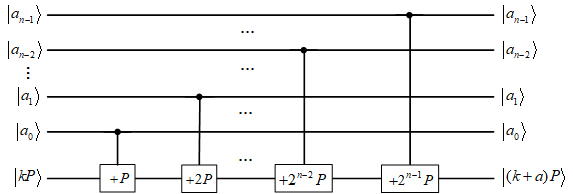}
	\caption{The general method to calculate $aP$ by quantum circuit.}
	\label{fig:aP}
\end{figure}

\begin{figure}[H]
    \centering
	\includegraphics[scale=0.4]{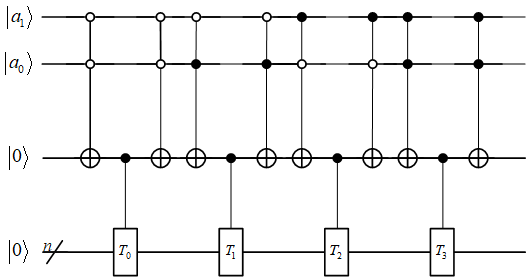}
 \qquad
 	\includegraphics[scale=0.4]{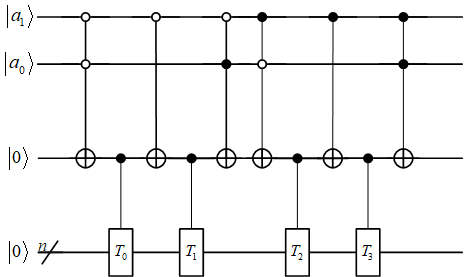}
    \caption{Quantum circuits of windowed arithmetic at $m=2$. Circuit of right one is the simplified of left one. }
    \label{fig:win-2}
\end{figure}

The general method to calculate $aP$ by quantum circuit is to expand $a$ binary and control the operation of $P$ by using each bit of $a$ respectively, i.e.,
	\begin{align*}
		aP=&(2^{n-1}a_{n-1}+2^{n-2}a_{n-2}+...+2a_{1}+a_0)P\\
		=&2^{n-1}a_{n-1}P+2^{n-2}a_{n-2}P+...+2a_{1}P+a_0P,
	\end{align*}
the circuit is shown in Fig.~\ref{fig:aP}.
The circuit designed by this method is the point addition operation of $n$-controlled. In Ref.~\cite{haner2020improved}, it is pointed out that $m$ $a_i$ can be selected first and the $2^m$ cases $a'P$ represented by $m$ $a_i$ can be calculated and stored in an $n$-qubit register, where $a'=\sum_{j=1}^{m}2^{i_j}a_{i_j}$.Then the $a'P$ is used to perform the point addition operation on the group of elliptic curves. The left circuit in Fig.~\ref{fig:win-2} shows the situation when $m=2$, where $T_i$ represents four cases of $a'$ respectively. Only the abscissa of point $P(x,y)$ is shown in the figure and $\frac{n}{2}$ CNOTs are required on average. Therefore, it is estimated that a total of $8$ Toffoli and $4n$ CNOTs are needed for the calculation point $P(x,y)$, i.e., $(4n+48)$ CNOTs.

In the case of $m$, $2^{m+1}$ m-controlled CNOTs (i.e., $2^{m+1}(2m-3)$ Toffoli gates) and $2^mn$ CNOTs are required in the circuit, so a total of $(24m+n-36)\cdot2^m$ CNOTs are required. However, we improve the circuit above to the right one in Fig.~\ref{fig:win-2}, thus we just need $(4n+26)$ CNOTs when $m=2$.

\begin{figure}[H]
	\centering
	\includegraphics[scale=0.5]{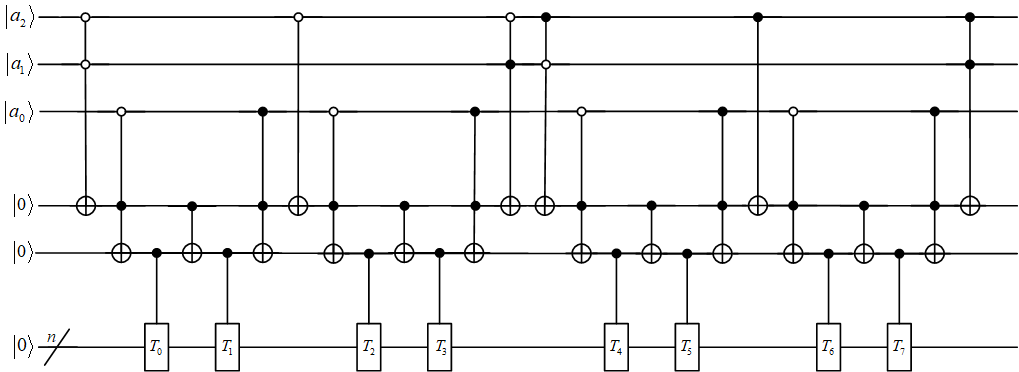}
	\caption{The simplified circuit at $m=3$.}
	\label{fig:win-m-3}
\end{figure}

\begin{figure}[H]
	\centering
	\includegraphics[scale=0.3]{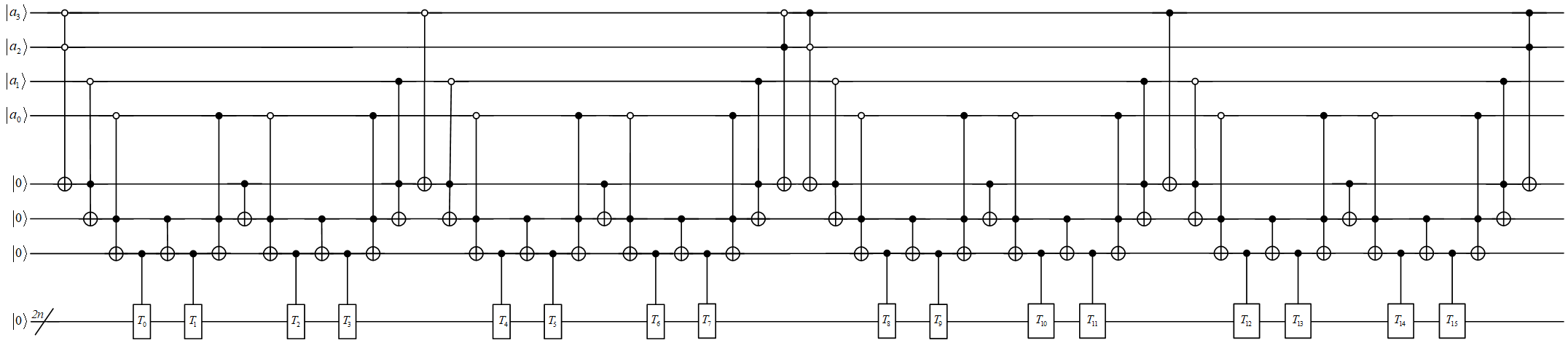}
	\caption{The simplified circuit at $m=4$.}
	\label{fig:win-m-4}
\end{figure}

The Fig.~\ref{fig:win-m-3} and Fig.~\ref{fig:win-m-4} are the situation when $m=3$ and $m=4$, respectively. According to the recursive formula, $(2^{m+1}-4)$ Toffoli and $[(n+1)\cdot 2^m-2]$ are required for $m$ with $m\geq3$. Thus the total number of CNOT required can be reduced to $[(n+13)\cdot 2^m-26]$.


\section*{Quantum circuits of point addition on elliptic curves groups}
\label{sec: QC for ecdlp}
Before we describe the construction of basic arithmetic used in point addition on the elliptic curves groups. In this section, we design a new algorithm to calculate point addition reversibly out place (storing the results in a new register) which reduces the number of CNOT gate of modular inversion and modular multiplication compared to the in place(replacing the input value by the sum) given by Ref.~\cite{roetteler2017quantum}, while using $O(n^2)$ qubits. Based on new approach for point addition, this section gives the schematic circuit of overall extended Shor's algorithm for ECDLP and then applies windowed arithmetic~\cite{gidney2019windowed} to obtain the windowed scalar multiplication of the given point on elliptic curves.
\subsection*{Controlled form of point addition}
Controlled form of point addition on elliptic curve, this algorithm operates on a quantum register holding the point $P_1=(x_1,y_1)\neq\emptyset$, a control bit $ctrl$ and ten auxiliary bits $c_i$. The second point $P_2=(x_2,y_2)\neq\emptyset, P_2\neq\pm P_1$ is assumed to be precomputed classical constant. If $ctrl=1$, the algorithm correctly calculates $c_9\leftarrow x_1+x_2,c_{10}\leftarrow y_1+y_2$; if $ctrl=0, c_9\leftarrow x_1,c_{10}\leftarrow y_1$.

Table \ref{tab:P1+P2} and Table \ref{tab:restore} describe the process of calculating $P_1+P_2$ and restoring auxiliary bits, respectively.

\begin{figure}[H]
	\centering
	\includegraphics[scale=0.45]{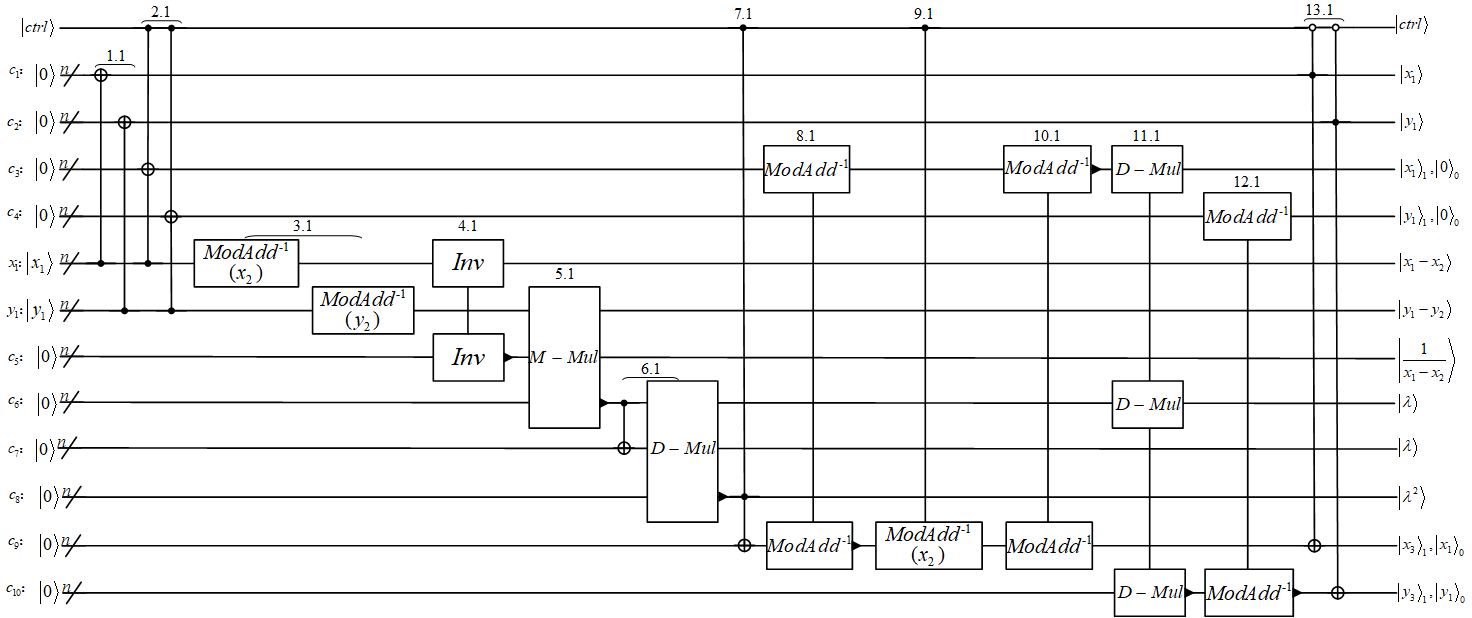}
	\caption{Controlled point addition to calculate $(P_1+P_2)\mod p$.}
	\label{fig:ctrl-pointadd-L}
\end{figure}

\begin{figure}[H]
	\centering
	\includegraphics[scale=0.4]{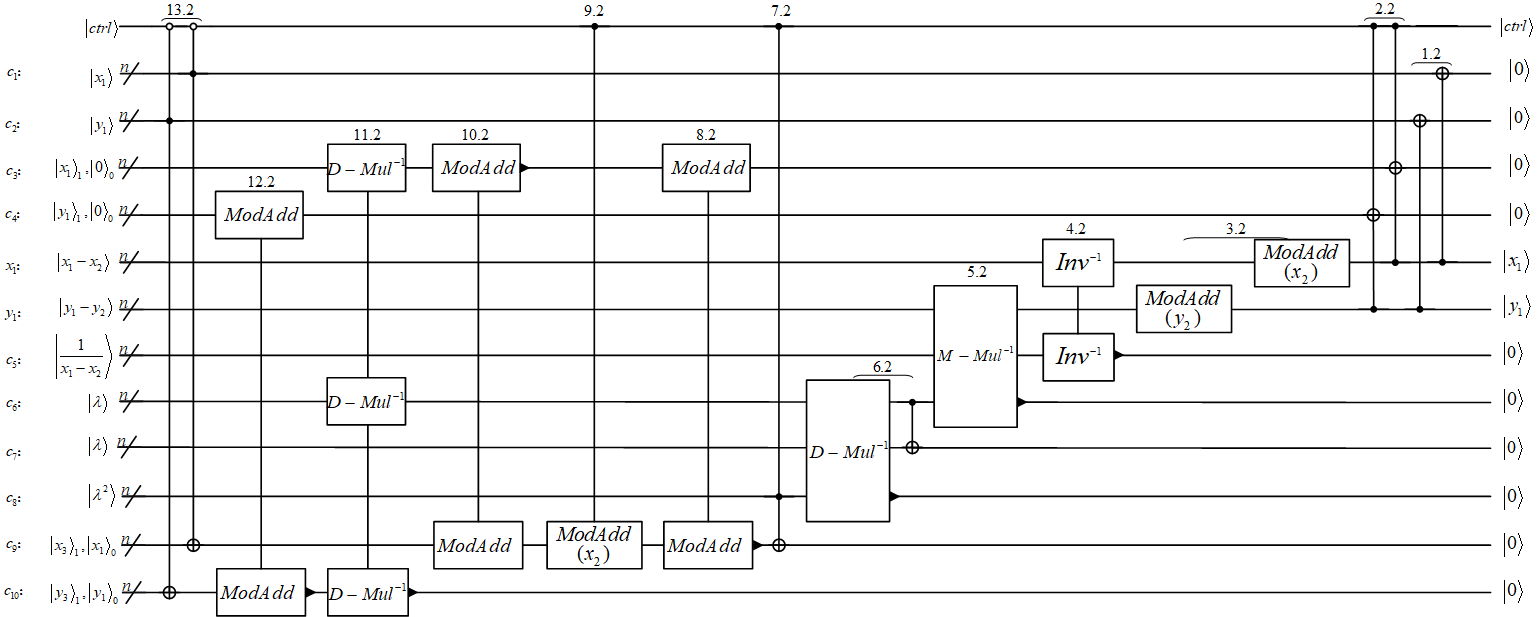}
	\caption{The inverse operation of controlled point addition to restore the anxiliary bits.}
	\label{fig:ctrl-pointadd-R}
\end{figure}

\begin{table}[H]
	\caption{The steps from $(x_1,y_1)$ to $(x_3,y_3)$ by point addition. Symbols $|\cdot\rangle_1$ and $|\cdot\rangle_0$ respectively represent the state when the control bit is $1$ and $0$. The state in the table 1 represents the change of the quantum register corresponding to each step and the unwritten states are the same as the states in the previous step. }
		\centering
		\begin{tabular}{ccc}
			\hline
			&process& the change in value  \\
			\hline
			1.1& $CNOT\quad c_1,c_2,x_1,y_1$ &$c_1\leftarrow x_1;c_2\leftarrow y_1$\\
			2.1& $Ctrl-CNOT\quad c_3,c_4,x_1,y_1,ctrl$ &$c_3\leftarrow [x_1]_1,[0]_0;c_4\leftarrow [y_1]_1,[0]_0$\\			
			3.1 &$ModAdd^{-1}(\cdot)\quad x_1,y_1,x_2,y_2$& $x_1\leftarrow x_1-x_2;y_1\leftarrow y_1-y_2$\\
			4.1 &$Inv\quad c_5,x_1$ &$c_5\leftarrow \frac{1}{x_1-x_2}$\\
			5.1 &$M-Mul\quad c_6,y_1,c_5$& $c_6\leftarrow\lambda$\\	
			6.1 &$D-Mul \quad c_8,c_6,c_7$ &$c_7\leftarrow \lambda;c_8\leftarrow \lambda^2$\\
			7.1 &$Ctrl-CNOT\quad c_9,c_8,ctrl$& $c_9\leftarrow [\lambda^2]_1,[0]_0$\\
			8.1 &$ModAdd^{-1}\quad c_9,c_3$ &$c_9\leftarrow [\lambda^2-x_1]_1,[0]_0$\\
			9.1 &$Ctrl-ModAdd^{-1}(\cdot)\quad c_9,x_2,ctrl$& $c_9\leftarrow [\lambda^2-x_1-x_2=x_3]_1,[0]_0$\\
			10.1 &$ModAdd^{-1}\quad c_3,c_4$ &$c_3\leftarrow [x_1-x_3]_1,[0]_0$\\
			11.1 &$D-Mul\quad c_{10},c_3,c_7$& $c_{10}\leftarrow [\lambda(x_1-x_3)]_1,[0]_0$\\	
			12.1 &$ModAdd^{-1}\quad c_{10},c_4$& $c_{10}\leftarrow [\lambda(x_1-x_3)-y_1=y_3]_1,[0]_0$\\
			13.1 &$Ctrl-CNOT\quad c_9,c_{10},c_1,c_2,ctrl$& $c_9\leftarrow [x_3]_1,[x_1]_0;c_{10}\leftarrow [y_3]_1,[y_1]_0$\\				 
			\hline
	\end{tabular}
	\label{tab:P1+P2}
\end{table}
\begin{table}[H]
	\centering
	\caption{The steps to restore the anxiliary bits. Symbols $|\cdot\rangle_1$ and $|\cdot\rangle_1$ respectively represent the state when the control bit is $1$ and $0$. The state in the table represents the change of the quantum register corresponding to each step and the unwritten states are the same as the states in the previous step. }
		\centering
		\begin{tabular}{ccc}
			\hline
			&process& the change in value  \\
			\hline
			13.2 &$Ctrl-CNOT\quad c_9,c_{10},c_1,c_2,ctrl$& $c_9\leftarrow [x_3]_1,[0]_0;c_{10}\leftarrow [y_3]_1,[0]_0$\\	
			12.2 &$ModAdd\quad c_{10},c_4$& $c_{10}\leftarrow [\lambda(x_1-x_3)]_1,[0]_0$\\			
			11.2 &$D-Mul^{-1}\quad c_{10},c_3,c_7$& $c_{10}\leftarrow 0$\\			
			10.2 &$ModAdd\quad c_3,c_4$ &$c_3\leftarrow [x_1]_1,[0]_0$\\
			9.2 &$Ctrl-ModAdd(\cdot)\quad c_9,x_2,ctrl$& $c_9\leftarrow [\lambda^2-x_1]_1,[0]_0$\\
			8.2 &$ModAdd\quad c_9,c_3$ &$c_9\leftarrow [\lambda^2]_1,[0]_0$\\									7.2 &$Ctrl-CNOT\quad c_9,c_8,ctrl$& $c_9\leftarrow 0$\\
			6.2 &$D-Mul^{-1} \quad c_8,c_6,c_7$ &$c_7\leftarrow 0;c_8\leftarrow 0$\\
			5.2 &$M-Mul^{-1}\quad c_6,y_1,c_5$& $c_6\leftarrow 0$\\	
			4.2 &$Inv^{-1}\quad c_5,x_1$ &$c_5\leftarrow 0$\\
			3.2 &$ModAdd(\cdot)\quad x_1,y_1,x_2,y_2$& $x_1\leftarrow x_1;y_1\leftarrow y_1$\\
			2.2& $Ctrl-CNOT\quad c_3,c_4,x_1,y_1,ctrl$ &$c_3\leftarrow 0;c_4\leftarrow 0$\\			
			1.2& $CNOT\quad c_1,c_2,x_1,y_1$ &$c_1\leftarrow 0;c_2\leftarrow 0$\\	 
			\hline
	\end{tabular}
	\label{tab:restore}
\end{table}
Fig.~\ref{fig:ctrl-pointadd-L} and Fig.~\ref{fig:ctrl-pointadd-R} show quantum circuits corresponding to Table \ref{tab:P1+P2} and Table \ref{tab:restore}. The quantum registers all consist of $n$ logical qubits, whereas $|ctrl\rangle$ is a single logical qubits. Thus the number of CNOT is $896n^2+1064n+14$.

\begin{figure}[H]
	\centering
	\includegraphics[scale=0.35]{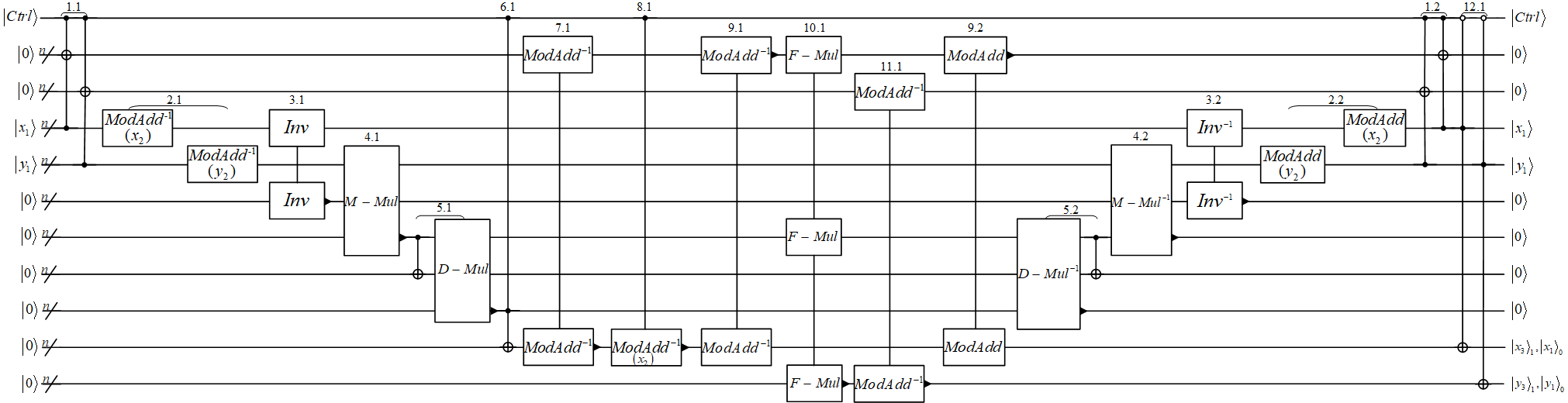}
	\caption{The last calculation of controlled point addition.}
	\label{fig:point-addition}
\end{figure}

After each calculation of $P_1+P_2$, the result will be used as the next input $P_1$ for a new calculation and then will be restored as an auxiliary bit. However, the result of the last calculation should be kept in the auxiliary register without any need to be restored. Therefore, the circuit of the last calculation is modified as shown in Fig.~\ref{fig:point-addition}. And the number of CNOT is $886n^2+783.5n-18.5$.

Therefore, the schematic quantum circuit of overall extended Shor's algorithm for ECDLP can be obtained by combining the Fig.~\ref{fig:ctrl-pointadd-L} to Fig.~\ref{fig:point-addition}.
\begin{figure}[H]
	\centering
	\includegraphics[scale=0.4]{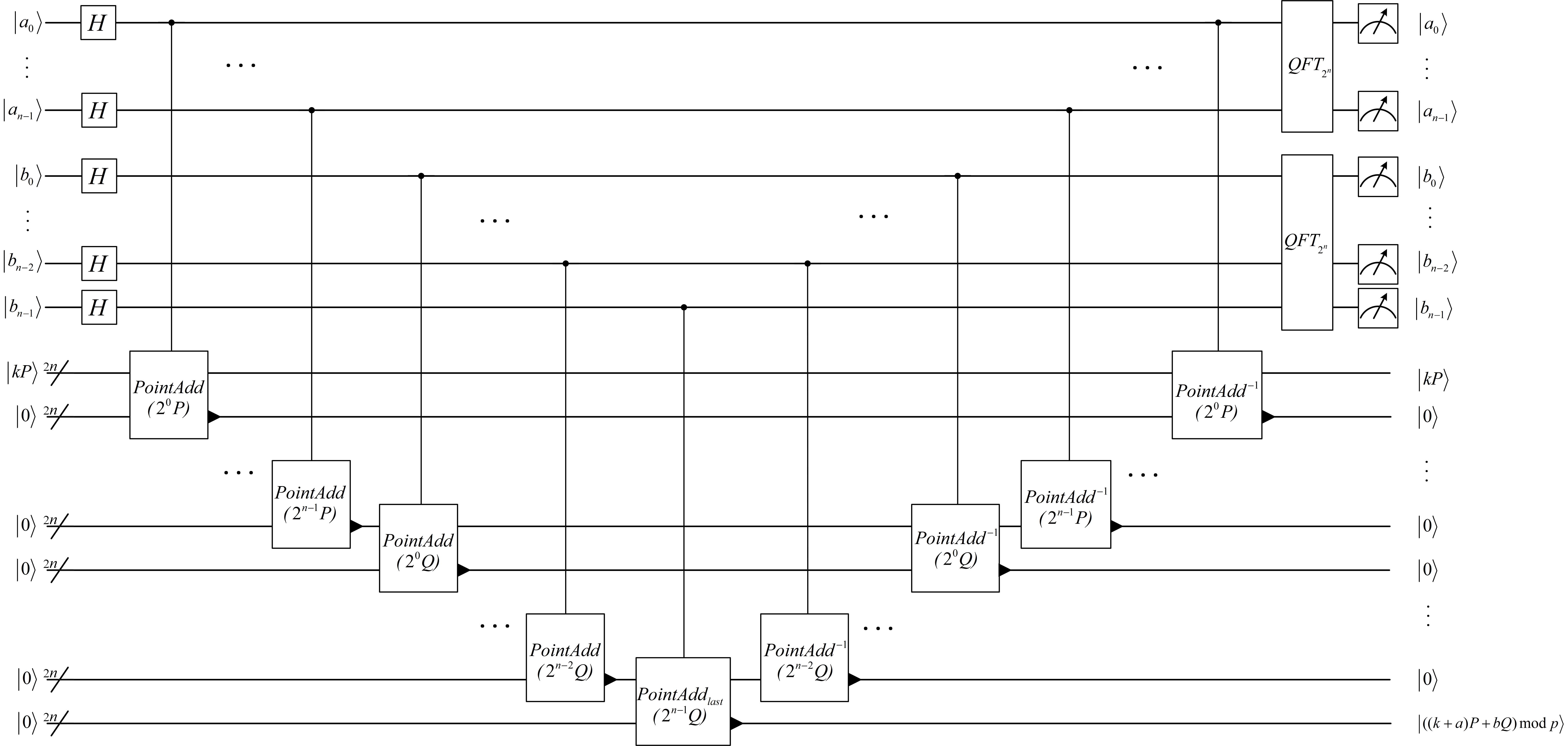}
	\caption{Schematic quantum circuit of overall extended Shor's algorithm for ECDLP.}
	\label{fig:ECDLP}
\end{figure}

\subsection*{Windowed form of point addition}
Windowed form of point addition on elliptic curve, this algorithm operates on a quantum register holding the point $P_1=(x_1,y_1)\neq\emptyset$, $P_2(x_2,y_2)\neq\emptyset,P_2\neq\pm P_1$ and eight auxiliary bits. In this form, the second point $P_2$ is stored in the quantum register as a quantum state and cannot be precomputed as a classical constant. 

Table~\ref{tab:win} and Table~\ref{tab:win-restore} describe the process of calculating $P_1+P_2$ by windowed arithmetic and restoring auxiliary bits, respectively.

\begin{table}[H]
	\centering
	\caption{The steps from $(x_1,y_1)$ to $(x_3,y_3)$ using windowed arithmetic by point addition. The state in the table 1 represents the change of the quantum register corresponding to each step and the unwritten states are the same as the states in the previous step.}
		\centering
		\begin{tabular}{ccc}
			\hline
			&process& the change in value  \\
			\hline
			1.1& $CNOT\quad w_1,w_2,x_1,y_1$ &$w_1\leftarrow x_1,w_2\leftarrow y_1$\\
			2.1 &$ModAdd^{-1}\quad x_1,y_1,x_2,y_2$& $x_1\leftarrow x_1-x_2,y_1\leftarrow y_1-y_2$\\
			3.1 &$Inv\quad w_3,x_1$ &$w_3\leftarrow \frac{1}{x_1-x_2}$\\
			4.1 &$M-Mul\quad w_4,y_1,w_3$& $w_4\leftarrow\lambda$\\	
			5.1 &$D-Mul \quad w_6,w_4,w_5$ &$w_6\leftarrow \lambda^2$\\
			6.1 &$CNOT\quad w_7,w_6$& $w_7\leftarrow \lambda^2$\\
			7.1 &$ModAdd^{-1}\quad w_7,w_1$ &$w_7\leftarrow \lambda^2-x_1$\\
			8.1 &$ModAdd^{-1}\quad w_7,x_2$& $w_7\leftarrow \lambda^2-x_1-x_2=x_3$\\
			9.1 &$ModAdd^{-1}\quad w_1,w_7$ &$w_1\leftarrow x_1-x_3$\\
			10.1 &$D-Mul\quad w_8,w_1,w_4$& $w_8\leftarrow \lambda(x_1-x_3)$\\	
			11.1 &$ModAdd^{-1}\quad w_8,w_2$& $w_8\leftarrow \lambda(x_1-x_3)-y_1=y_3$\\	 
			\hline
	\end{tabular}
	\label{tab:win}
\end{table}

\begin{figure}[H]
	\centering
	\includegraphics[scale=0.5]{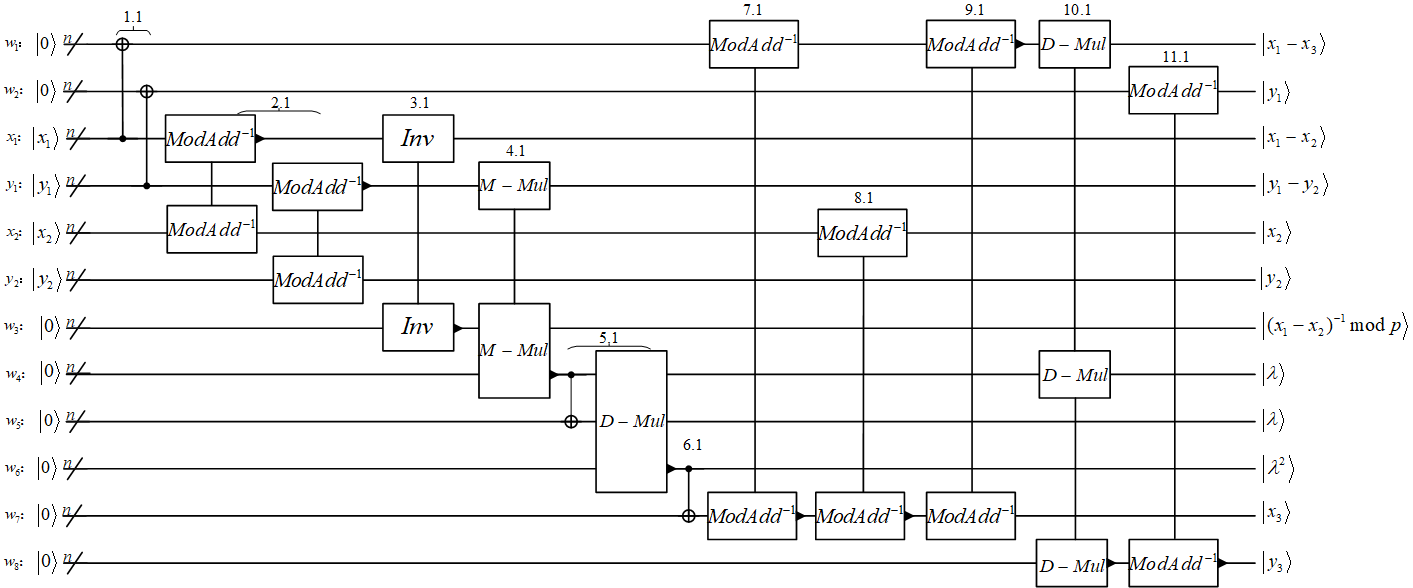}
	\caption{Windowed point addition.}
	\label{fig:window-pointadd}
\end{figure}

\begin{figure}[H]
	\centering
	\includegraphics[scale=0.45]{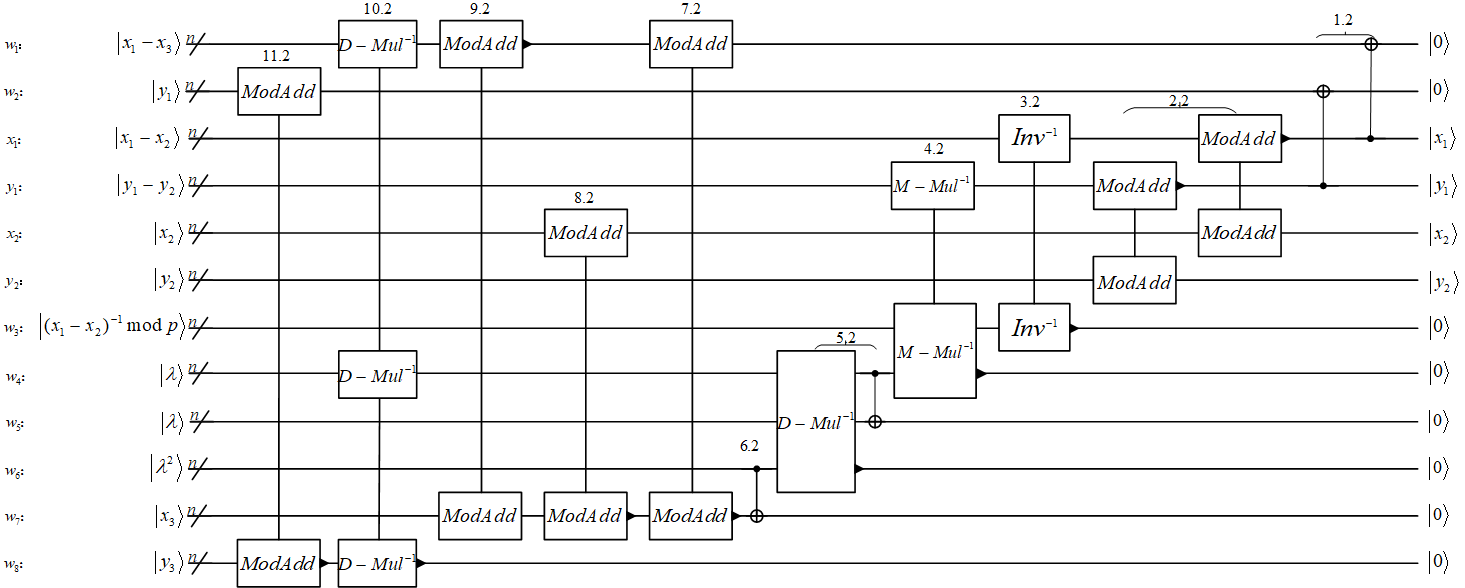}
	\caption{The inverse operation of windowed point addition to restore the anxiliary bits.}
	\label{fig:inv-window-pointadd}
\end{figure}

\begin{table}[H]
	\centering
	\caption{The steps to restore the anxiliary bits. The state in the table represents the change of the quantum register corresponding to each step and the unwritten states are the same as the states in the previous step. }
		\centering
		\begin{tabular}{ccc}
			\hline
			&process& the change in value  \\
			\hline	
			11.2& $ModAdd\quad w_8,w_2$& $w_8\leftarrow \lambda(x_1-x_3)$\\	 
			10.2& $D-Mul^{-1}\quad w_8,w_1,w_4$& $w_8\leftarrow0$\\	
			9.2& $ModAdd\quad w_1,w_7$ &$w_1\leftarrow x_1$\\
			8.2& $ModAdd\quad w_7,x_2$& $w_7\leftarrow \lambda^2-x_1$\\	
			7.2& $ModAdd\quad w_7,w_1$ &$w_7\leftarrow\lambda^2$\\
			6.2& $CNOT\quad w_7,w_6$& $w_7\leftarrow 0$\\
			5.2& $D-Mul^{-1} \quad w_6,w_4,w_5$ &$w_6\leftarrow 0$\\
			4.2& $M-Mul^{-1}\quad w_4,y_1,w_3$& $w_4\leftarrow0$\\	
			3.2& $Inv^{-1}\quad w_3,x_1$ &$w_3\leftarrow 0$\\
			2.2& $ModAdd\quad x_1,y_1,x_2,y_2$& $x_1\leftarrow x_1,y_1\leftarrow y_1$\\
			1.2& $CNOT\quad w_1,w_2,x_1,y_1$ &$w_1\leftarrow 0,w_2\leftarrow 0$\\					
			\hline
	\end{tabular}
	\label{tab:win-restore}
\end{table}
Fig.~\ref{fig:window-pointadd} and Fig.~\ref{fig:inv-window-pointadd} show quantum circuits corresponding to Table~\ref{tab:win} and Table~\ref{tab:win-restore}. The quantum registers all consist of $n$ logical qubits. Thus the number of CNOT is $896n^2+1108n+36$.

\begin{figure}[H]
	\centering
	\includegraphics[scale=0.35]{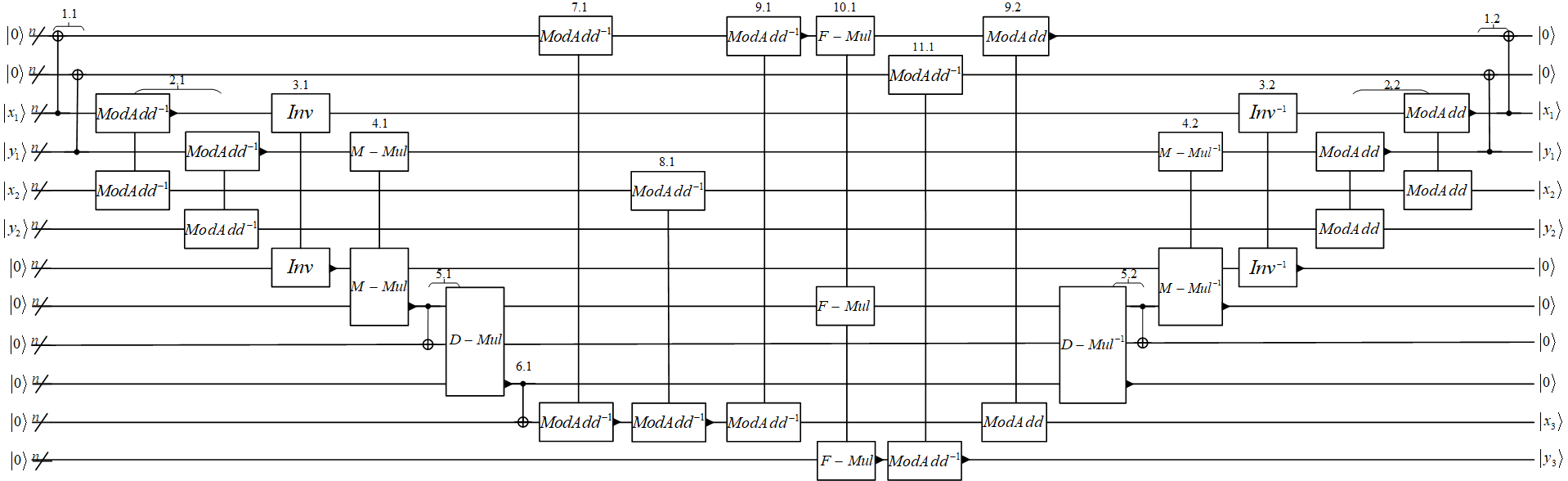}
	\caption{The last full round of windowed point addition.}
	\label{fig:last-window-pointadd}
\end{figure}

After each calculation of $P_1+P_2$, the result will be used as the next input $P_1$ for a new calculation and then will be restored as an auxiliary bit. However, the result of the last calculation should be kept in the auxiliary register without any need to be restored and the coefficients of $P$ and $Q$ are different in the extended Shor's quantum algorithm, it cannot be mixed in the application of using window arithmetic. Therefore, the circuit of the last calculation is modified as shown in Fig.~\ref{fig:last-window-pointadd} with $886n^2+833.5n+9.5$.

\begin{figure}[H]
	\centering
	\includegraphics[scale=0.4]{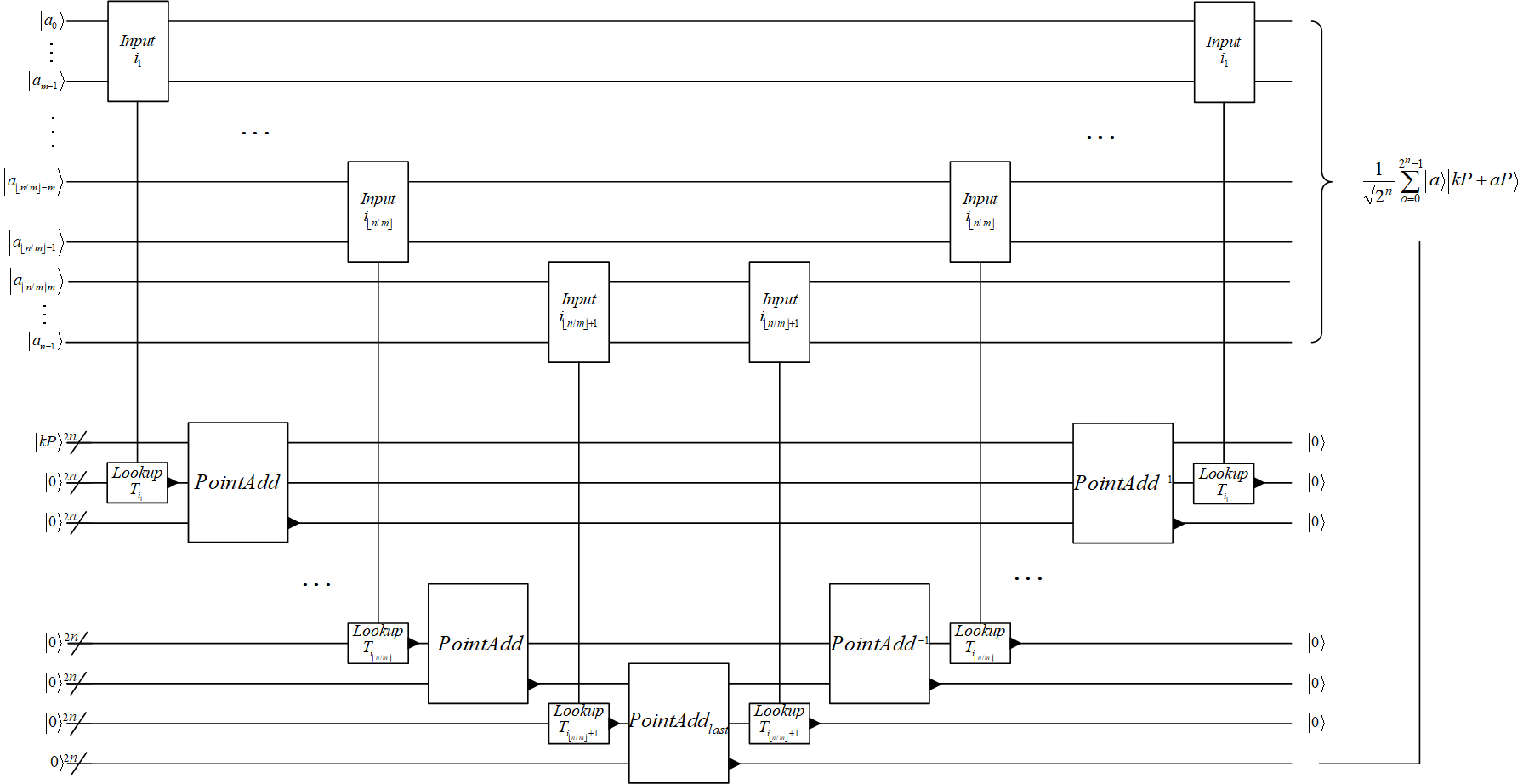}
	\caption{Schematic quantum circuit of overall extended Shor's algorithm for ECDLP using windowed arithmetic.}
	\label{fig:ECDLP-window}
\end{figure}

Fig.~\ref{fig:ECDLP-window} is the schematic quantum circuit to calculate ECDLP by the extended Shor's algorithm using windowed arithmetic, where the $Lookup$ is situation where several controlled operations can be merged into a single operation acting on a value produced by a small QROM lookup~\cite{gidney2019windowed} and the point addition is the circuit introduced in Fig.~\ref{fig:window-pointadd} to Fig.~\ref{fig:ECDLP-window}.

According to the process of calculating point addition, the number of CNOT gates for the first $2n-1$ point addition is $896n^2+1064n+14$ (including the circuits for recovering auxiliary bits) and the $2n$-th has $886n^3+783.5n-18.5$ CNOTs. Therefore, the number of CNOT needed to calculate the point addition of ECDLP using controlled point addition is $1792n^3+2118n^2-252.5n-32.5$. When using the form of window, modular subtraction of a constant is changed to $ModAdd^{-1}$ and the CNOT number increases from $43n-2.5$ to $61n+6$. At the same time, the number of CNOT in the controlled circuit increases from $46n+11$ to $71n+17$, thus the number of CNOT gates for the first $n-1$ point addition is $896n^2+1108n+36$ and the $n$-th has $886n^3+833.5n+9.5$ CNOTs in calculating $(a+k)P\mod p$. so the number of CNOT needed to calculate the point addition of ECDLP using the form of Window is $2\lceil\frac{n}{m}\rceil[(n+13)\cdot 2^{m+1}+896n^2+1108n+4]-20n^2-549n-53$. 

Now we analyze the whole circuit of the extended Shor's algorithm to obtain a specific CNOT number. It can be inferred from the form of the formula that only when $m=O(\log n)$, $N$ is a polynomial function of $n$. We calculate $\frac{\partial N(n, m)}{\partial m}$, and for each $n_i \in (128, 521)$, we use Matlab to approximate the zero $m_i$ of $\frac{\partial N(n_i, m)}{\partial m}$ to obtain a pair $(n_i, m_i)$. Because $m$ should be an integer, we round each $m_i$ up and down to get $m_i'$ and $m_i''$, respectively. Then letting $N_{i_{\rm min}} = \min(N(n_i, m_i'), N(n_i, m_i''))$ for each $i$ and fitting $N$ with respect to $n$ based on all the pairs $(n_i, N_{i_{\rm min}})$, we obtain $N= 1237n ^3/\log n$. Combine the $2n^2+n$ CNOT gates used for two $QFT_n$, then there has the total number of CNOT of the extended Shor's algorithm for ECDLP is $N= 1237n ^3/\log n+2n^2+n$. Ref.~\cite{yang2013post} gives the lower limit of time for executing a CNOT gate in an ion trap quantum computer, which is about $2.85\times10^{-4}s$. Combined with the number of CNOT to run the extend Shor's algorithm, the time to break $512$-bit ECDLP takes at least $51$ years after three levels of coding.

\section*{Discusion and conclusion}
\label{sec: dis}
Although there have been many attempts to improve the qubit number or the circuit depth of the extended Shor’s algorithm for ECDLP, their focus has not been on optimizing the number of CNOT gates, which greatly affects the time to run the algorithm in an ion trap quantum computer. In this paper, we improve the quantum circuit of basic arithmetic operations, including modulus subtraction, three different modulus multiplication, modulus inverse and windowed arithmetic. Table~\ref{tab:basic cnot} summarizes the CNOT numbers for the basic arithmetic. We further improve the quantum citcuit of extended Shor's algorithm. Based on this work, the quantum circuit that run the extended Shor's algorithm is greatly improved. We reduce the CNOT gates number from $O(n^3)$ to $O(n^3/\log n)$, and calculate the specific number of CNOT gates required by the algorithm. The time required by the extended Shor's algorithm to attack $512$-bit ECDLP are estimated, i.e., $51$ years, which means it is hard to attack ECDLP using ion trap quantum computer in a reasonable time. However, this estimated time does not take into account the fault tolerance of the circuit, which we will study in the future.  

According to the results of CNOT number, we can consider the lower bound of CNOT number required by the extended Shor's algorithm. If we assume that the time required to run extended Shor's algorithm is $T$, the time required to execute a CNOT is $t$, and the lower bound of the number of CNOTs is $N$, which is a function of the number of qubits $n$. Thus, the lower bound of the run time of the extended Shor's algorithm can be expressed as $T=N(n)t$. Modular inverse can be constructed using the basic arithmetic operations, such as modular addition. Therefore, the number of CNOTs required for modular inverse must be greater than that required for modular addition. Because the quantum circuit of modular addition adds a modular operation, the number of CNOTs required by modular addition must be larger than that of the addition circuit. For two $n$ qubits $x, y$, we have that $c_{i+1}=x_i+(x_i+y_i)(x_i+c_i), s_i=x_i+y_i+c_i$, where $x_i$ and $y_i$ are the $i$-$th$ bits of the binary representation of $x,y$, $c_{i+1}$ is the $i$-$th$ carry, and $s_i$ is the sum of the $i$-$th$ bits. Therefore, each qubit addition requires at least one Toffoli and three CNOTs. Thus, the addition of $n$ qubits requires at least $9n$ CNOTs. However, we have not obtain a specific lower bound of the number of CNOT gates required to run the extended Shor's algorithm, so we plan to calculate the lower bound more tight in our next work.

\begin{table}[H]
\scriptsize
\centering
\caption{The CNOT numbers for the basic arithmetic.}
  \begin{tabular}{cccc}
    \hline
    The basic arithmetic & Number of Toffoli & Number of CNOT & Total number of CNOT \\ \hline
    1-Add$_y$ (unknown $y$) & $2n$ & $4n+1$ & $16n+1$\\
    ctrl-1-Add$_y$ (unknown $y$) & $4n+1$ &$2n$ & $26n+6$\\
    2-Add$_y$ (know $y$) & $2n-1$ & $2n+0.5$ & $14n-5.5$\\
    ctrl-2-Add$_y$ & $2n$ &$5n+1$ & $17n+1$\\
    1-comp & $2n$ & $1$ & $12n+1$\\
    2-comp$_y$ (know $y$) & $2n$ &$1$ & $12n+1$\\
    ctrl-2-comp$_y$ (know $y$) & $2n+1$ &$1$ & $12n+7$\\
    2-comp$_y$ (unknown $y$) & $2n$ &$4n+1$ & $16n+1$\\
    ctrl-2-comp$_y$ (unknown $y$) & $2n+1$ &$4n+1$ & $16n+7$\\
    Mod-Sub$_y$ or Mod-Add$_y$ (know $y$) & $6n-1$ &$7n+3.5$ & $43n-2.5$\\	
    ctrl-Mod-Sub$_y$ (known $y$) & $6n+1$ &$10n+5$ & $46n+11$  \\  		
    Mod-Add$_y$ (unknown $y$) & $8n$ &$13n+6$ & $61n+6$\\  
    ctrl-Mod-Add$_y$ (unknown $y$) & $10n+2$ &$11n+5$ & $71n+17$\\            
    neg & $2n-1$ &$2n-0.5$ & $14n-5.5$\\    
    ctrl-neg & $2n$ &$6n+1$ & $18n+1$\\
    1-shift & - &$2n$ & $2n$\\ 
    ctrl-1-shift & $2n$ &- & $12n$\\ 
    2-shift & - &$3n$ & $3n$\\ 
    ctrl-2-shift & $n$ &$2n$ & $8n$\\ 
    Shift-Mod & $4n$&$7n+3$& $31n+15$\\
    M-MM (half)& $6n^2+5n-1$ &$9n^2+9n+0.5$ & $45n^2+39n-4.5$\\
    D-Mul (half)& $8n^2$ &$9n^2+1.5$ & $57n^2+2.5n$\\ \hline
  \end{tabular}
  \label{tab:basic cnot}
\end{table}

\bibliographystyle{apsrev4-1}
\bibliography{bib}

\section*{Appendix}

We first prove in detail that whether the input state of the third quantum register is $|0\rangle$ or $|1\rangle$ has no effect on the measurement probability. Then, we give the specific derivation process of the number of CNOT gates in the n-controlled-NOT.
\subsection*{The value of the input state has no effect on the result}\label{appen:proof}

    \begin{proof}
$(1)$ When the input state is $|0\rangle$, we have
\begin{eqnarray}\nonumber
|0\rangle|0\rangle|0\rangle&\rightarrow&\frac{1}{p-1}\sum_{a_1=0}^{p-2}\sum_{b_1=0}^{p-2}|a_1\rangle|b_1\rangle |0\rangle\nonumber\\
&\rightarrow&\frac{1}{p-1}\sum_{a_1=0}^{p-2}\sum_{b_1=0}^{p-2}|a_1\rangle|b_1\rangle |(a_1P+b_1Q)\mod p\rangle\nonumber\\
&\rightarrow&\frac{1}{(p-1)q}\sum_{a_1,b_1=0}^{p-2}\sum_{c_1,d_1=0}^{q-1}\exp[\frac{2\pi i}{q}(a_1c_1+b_1d_1)]\nonumber\\
&&\qquad\qquad\qquad\qquad|c_1\rangle|d_1\rangle |(a_1P+b_1Q)\mod p\rangle\nonumber\\
&=&\frac{1}{(p-1)q}\sum_{a_1,b_1=0}^{p-2}\sum_{c_1,d_1=0}^{q-1}\exp[\frac{2\pi i}{q}(a_1c_1+b_1d_1)]\nonumber\\
&&\qquad\qquad\qquad\qquad|c_1\rangle|d_1\rangle |(a_1+b_1m)P\mod p\rangle\nonumber,
\end{eqnarray}
where $a_1+b_1m=l_1\mod(p-1)$, so $a_1=l_1-b_1m-(p-1)\lfloor\frac{l_1-b_1m}{p-1}\rfloor$ , thus the probability of getting a $|c\rangle|d\rangle|y\rangle$ is
\[\left|\frac{1}{(p-1)1}\sum_{a_1,b_1=0}^{p-2}\exp[\frac{2\pi i}{q}(a_1c_1+b_1d_1)]\right|^2\] 
\[=\left|\frac{1}{(p-1)q}\sum_{b_1=0}^{p-2}\exp[\frac{2\pi i}{q}(l_1c_1-b_1c_1m+b_1d_1-(p-1)c_1\lfloor\frac{l_1-b_1m}{p-1}\rfloor)]\right|^2\]
so when taking all of $|y\rangle $, the probability of $|c\rangle |d\rangle $ is 
\begin{eqnarray}
	\sum_{l_1=0}^{p-2}|\frac{1}{(p-1)q}\sum_{b_1=0}^{p-2}\exp[\frac{2\pi i}{q}(l_1c_1-b_1c_1m+b_1d_1-(p-1)c_1\lfloor\frac{l_1-b_1m}{p-1}\rfloor)]|^2
\end{eqnarray}

$(2)$ When the input state is $|kP\rangle$, we have
\begin{eqnarray}\nonumber
|0\rangle|0\rangle|kP\rangle&\rightarrow&\frac{1}{p-1}\sum_{a_2=0}^{p-2}\sum_{b_2=0}^{p-2}|a_2\rangle|b_2\rangle |kP\rangle\nonumber\\
&\rightarrow&\frac{1}{p-1}\sum_{a_2=0}^{p-2}\sum_{b_2=0}^{p-2}|a_2\rangle|b_2\rangle |((a_2+k)P+b_2Q)\mod p\rangle\nonumber\\
&\rightarrow&\frac{1}{(p-1)q}\sum_{a_2,b_2=0}^{p-2}\sum_{c_2,d_2=0}^{q-1}\exp[\frac{2\pi i}{q}(a_2c_2+b_2d_2)]\nonumber\\
&&\qquad\qquad\qquad\qquad|c_2\rangle|d_2\rangle |((a_2+k)P+b_2Q)\mod p\rangle\nonumber\\
&=&\frac{1}{(p-1)q}\sum_{a_2,b_2=0}^{p-2}\sum_{c_2,d_2=0}^{q-1}\exp[\frac{2\pi i}{q}(a_2c_2+b_2d_2)]\nonumber\\
&&\qquad\qquad\qquad\qquad|c_2\rangle|d_2\rangle |((a_2+k)+b_2m)P\mod p\rangle\nonumber,
\end{eqnarray}
where $a_2+k+b_2m=l_2\mod(p-1)$, so $a_2=l_2-k-b_2m-(p-1)\lfloor\frac{l_2-k-b_2m}{p-1}\rfloor$, thus the probability of getting a $|c\rangle|d\rangle|y\rangle$ is
\[|\frac{1}{(p-1)q}\sum_{a_2,b_2=0}^{p-2}\exp[\frac{2\pi i}{q}(a_2c_2+b_2d_2)]|^2\] 
\[=|\frac{1}{(p-1)q}\sum_{b_2=0}^{p-2}\exp[\frac{2\pi i}{q}(l_2c_2-b_2c_2m-kc_2+b_2d_2-(p-1)c_2\lfloor\frac{l_2-k-b_2m}{p-1}\rfloor)]|^2\]
so when taking all of $|y\rangle $, the probability of $|c\rangle |d\rangle $ is 
\begin{eqnarray}
\sum_{l_2=0}^{p-2}|\frac{1}{(p-1)q}\sum_{b_2=0}^{p-2}\exp[\frac{2\pi i}{q}(l_2c_2-b_2c_2m-kc_2+b_2d_2-(p-1)c_2\lfloor\frac{l_2-k-b_2m}{p-1}\rfloor)]|^2
\end{eqnarray}

$(3)$ To prove that it doesn't matter whether the input is $|0\rangle$ or $|kP\rangle$, just prove that $(5)=(6)$\\
There has
\[\sum_{l_1=0}^{p-2}|\frac{1}{(p-1)q}\sum_{b_1=0}^{p-2}\exp[\frac{2\pi i}{q}(l_1c_1-b_1c_1m+b_1d_1-(p-1)c_1\lfloor\frac{l_1-b_1m}{p-1}\rfloor)]|^2\]
\begin{align*}
&=\sum_{l_1=0}^{p-2-k}|\frac{1}{(p-1)q}\sum_{b_1=0}^{p-2}\exp[\frac{2\pi i}{q}(l_1c_1-b_1c_1m+b_1d_1-(p-1)c_1\lfloor\frac{l_1-b_1m}{p-1}\rfloor)]|^2\\
&\quad+\sum_{l_1=p-1-k}^{p-2}|\frac{1}{(p-1)q}\sum_{b_1=0}^{p-2}\exp[\frac{2\pi i}{q}(l_1c_1-b_1c_1m+b_1d_1-(p-1)c_1\lfloor\frac{l_1-b_1m}{p-1}\rfloor)]|^2\\
&\\
&=\textcircled{1}+\textcircled{2};
\end{align*}
\[\sum_{l_2=0}^{p-2}|\frac{1}{(p-1)q}\sum_{b_2=0}^{p-2}\exp[\frac{2\pi i}{q}(l_2c_2-b_2c_2m-kc_2+b_2d_2-(p-1)c_2\lfloor\frac{l_2-k-b_2m}{p-1}\rfloor)]|^2\]
\begin{align*}
&=\sum_{l_2=0}^{k-1}|\frac{1}{(p-1)q}\sum_{b_2=0}^{p-2}\exp[\frac{2\pi i}{q}(l_2c_2-b_2c_2m-kc_2+b_2d_2-(p-1)c_2\lfloor\frac{l_2-k-b_2m}{p-1}\rfloor)]|^2\\
&\quad+\sum_{l_2=k}^{p-2}|\frac{1}{(p-1)q}\sum_{b_2=0}^{p-2}\exp[\frac{2\pi i}{q}(l_2c_2-b_2c_2m-kc_2+b_2d_2-(p-1)c_2\lfloor\frac{l_2-k-b_2m}{p-1}\rfloor)]|^2\\
\\
&=\textcircled{3}+\textcircled{4}.
\end{align*}
Because $c_1=c_2,d_1=d_2$, so we have $\textcircled{1}=\textcircled{4}, \textcircled{2}=\textcircled{3}$, then $(5)=(6)$. 
\end{proof}
\subsection*{The number of CNOT gates in n-controlled-NOT}
\begin{figure}[H]
	\centering
	\includegraphics[scale=0.4]{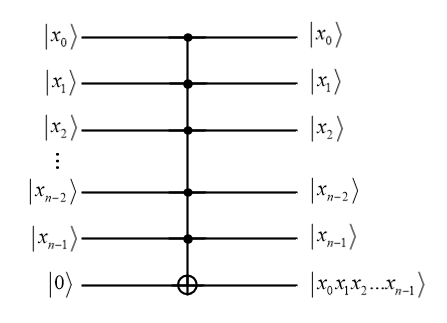}
	\caption{n-controlled-NOT}
	\label{fig:n-not-1}
\end{figure}

The Fig.~\ref{fig:n-not-1} is the quantum circuit of n-controlled-NOT,
which can be contrusted by $n-2$ auxiliary qubits and $2n-3$ Toffoli. The quantum circuit is as follows.
\begin{figure}[H]
	\centering
	\includegraphics[scale=0.3]{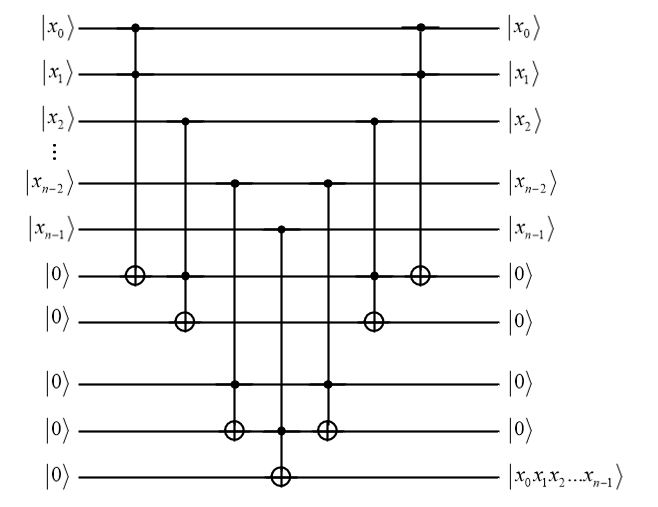}
	\caption{The equivalent form of n-controlled-NOT.}
	\label{fig:n-not-2}
\end{figure}

\end{document}